\documentclass[aps,prd,superscriptaddress,preprint,amsmath,amssymb]{revtex4-2}

\usepackage{graphicx}
\usepackage{dcolumn}
\usepackage{bm}
\usepackage{latexsym}
\usepackage{array}
\usepackage{setspace}

	\usepackage{amsthm}
	\theoremstyle{theorem}
	
    \usepackage[justification=centerlast]{caption}
    \usepackage[labelformat=simple]{subcaption}

\begin{document}
\title{Double-trace deformation in Keldysh field theory}

\author{Xiangyi~Meng}%
\email{xm@bu.edu}
\affiliation{Department of Physics, Boston University, Boston, Massachusetts 02215, USA}%
\affiliation{Department of Physics, Northeastern University, Boston, Massachusetts 02115, USA}%

\date{\today}

\begin{abstract}

The Keldysh formalism is capable of describing driven-dissipative dynamics of open quantum systems as nonunitary effective field theories that are not necessarily thermodynamical, thus often exhibiting new physics.
Here, we introduce a general Keldysh action that maximally obeys Weinbergian constraints, including locality, Poincar\'e invariance, and two ``\emph{CPT}'' constraints: complete positivity and trace preserving as well as charge, parity, and time reversal symmetry. We find that the perturbative Lindblad term responsible for driven-dissipative dynamics introduced therein has the natural form of a double-trace deformation $\mathcal{O}^2$, which, in the large $N$ limit, possibly leads to a new nonthermal conformal fixed point. 
This fixed point is IR when $\Delta<d/2$ or UV when $\Delta>d/2$ given $d$ the dimensions of spacetime and $\Delta$ the scaling dimension of $\mathcal{O}$. Such a UV fixed point being not forbidden by Weinbergian constraints may suggest its existence and even completion of itself, in contrast to the common sense that dissipation effects are always IR relevant. This observation implies that driven-dissipative dynamics is much richer than thermodynamics, differing in not only its noncompliance with thermodynamic symmetry (e.g., the fluctuation-dissipation relation) but its UV/IR relevance as well.
Examples including a $(0+1)$-$d$ harmonic oscillator under continuous measurement and a $(4-\epsilon)$-$d$ classic $O(N)$ vector model with quartic interactions are studied.

\end{abstract}


\maketitle


\section{Introduction}

Quantum mechanics, despite its extreme success in predicting and agreeing with every experimental outcome, still bemuses everyone by ``suspiciously'' postulating the existence of \emph{two} distinct but necessary time-evolution mechanisms, i.e., reversible unitary dynamics versus irreversible wave function collapse~\cite{q-mech}.
A better understanding of the origins and essentials of why a physical system may evolve in such two different ways seems the key to answering some of the most fundamental problems such as the black-hole information paradox~\cite{black-hole-inf-paradox_s18,black-hole-inf-paradox_mops15n,black-hole-inf-paradox_mops15b}, quantum measurement problem~\cite{q-collapse_blssu13,relativ-q-meas_mr02}, or even consistency of quantum mechanics itself~\cite{q-mech-contradict_fr18}. Not only that, it has recently been shown that the entanglement entropy for a quantum circuit undergoing the competition of unitary evolution and random measurements may behave quite distinctly by adjusting the rate of measurements~\cite{meas-induc_srn19}, a better understanding of which serves practical purposes such as universal quantum computing~\cite{q-comput_c09}, etc.

Even though the underlying physics remains unclear, the proper mathematical framework to integrate the two distinct time-evolution mechanisms has been well developed, such as the Choi-Kraus's theorem~\cite{ck-theor_c75,fundam-notion-q-theor} and the Gorini-Kossakowski-Sudarshan-Lindblad theorem~\cite{gksl-theor_gks76,gksl-theor_l76}. These theorems have further become the basis for the study of open quantum systems (OQS)~\cite{q-open-syst}, which aims to understand all possible forms of evolution of a quantum system that is not closed, and hence, unitarity can be lost. This does not only include examples such as how a system reaches equilibrium by being in contact with a \emph{thermal} reservoir (e.g., quantum Brownian motion~\cite{q-brown-motion_hpz92}), 
but also deals with a much more general scenario where, after tracing out part of the degrees of freedom (d.o.f.) of the system, the rest behaves effectively as an OQS coupled to an artificial environment consisting of the traced-out d.o.f. As for the latter scenario, the OQS undergoes driving force and dissipation, both induced by the environment, but is not necessarily thermalized. Led by driven-dissipative dynamics, the OQS may reach a \emph{nonthermal stationary state}---a realizable state of matter that can often be seen in many light-matter systems of Bose-Einstein condensates or Rydberg ensembles~\cite{keldysh_sbd16}.
Moreover, in the context of decoherence theory~\cite{decoherence_jz85,decoherence_z03}, the wave function collapse induced by quantum measurement~\cite{q-meas_cdgms10} may be framed as an environmental effect on the OQS as well, 
the idea of which has eventually triggered the development of a number of quantum collapse models~\cite{q-collapse_blssu13}, including the continuous spontaneous localization (CSL) model~\cite{contin-spontaneous-localis_gpr90}.

Unfortunately, for \emph{many-body} OQS, there are much fewer theoretical tools having been developed, despite the rich emergent and universal phenomena that the system may possess at the macroscopic scale. Most commonly, a field-theoretical tool is preferred, so that a number of well-developed techniques such as renormalization group (RG) 
will be applicable. A complete field-theoretical solution involves first the use of the Keldysh path integral~\cite{schwinger_s61,keldysh_k64,field-theor-non-equilib-syst}, which formulates how to set up a path integral that governs the evolution of not a wave function but a density matrix by \emph{doubling} the to-be-integrated field $\phi$ into two separate fields in the ket ($+$) and bra ($-$) basis, respectively~\cite{keldysh-i_hlr17,keldysh-ii_hlr17}. Next, tracing out the environment's d.o.f.~is carried out by the use of the Feynman-Vernon influential functional~\cite{feynman-vernon-path-integral_fv00}, leaving an effective field theory (EFT) as a Keldysh path-integral functional where only the bra and ket fields corresponding to the remaining d.o.f.~are kept.
This promising method of describing many-body OQS in a field-theoretical language has been actively discussed both in the nonrelativistic context~\cite{keldysh_sbd16} and relativistic context~\cite{keldysh-i_hlr17,keldysh-ii_hlr17} and has led to new research directions such as novel universality class in quantum phase transition~\cite{keldysh-new-univers_td14}, information loss in EFT~\cite{open-q-syst-inf-loss_b18,open-q-syst-renorm_abkl18}, etc. Note that a Feynman-Vernon influential functional is usually highly nonlocal. However, after applying the Born-Markov approximation~\cite{q-open-syst}, a Markovian form can often be produced where the time-evolution mechanisms are automatically local.

Inspired by this field-theoretical language, in this paper, 
we start directly with a general action formalized as a Keldysh path integral that works as a heuristic phenomenological model. Our propose is to study the driven-dissipative dynamics originating from mixing of the reversible and irreversible time-evolution mechanisms. We will discuss their implications rather than arguing about the origins of the mechanisms. Our Keldysh action is similar to the open-EFT Keldysh action introduced in Ref.~\cite{keldysh-renorm_bjlr17} but more general and only required to maximally obey ``Weinbergian'' constraints such as locality and Poincar\'e invariance~\cite{weinberg-q-field-theor}. Note that unitarity, however, must be revoked because of the existence of irreversible wave function collapse.

The main result we find for our Keldysh action is that, under appropriately renormalized perturbation, the nonunitary terms responsible for driven-dissipative dynamics have the natural form of a \emph{double-trace deformation}~\cite{deform-ads-cft_w02,deform_gk03,deform_a10,deform_gkp18} which has been very thoroughly studied---especially in the literature of conformal field theory (CFT)~\cite{conform-field-theor,conform-field-theor-3d}. In particular, in the large $N$ limit, it is argued that the existence of a double-trace deformation guarantees and produces a healthy RG flow from infrared (IR) backwards to ultraviolet (UV), suggesting that the theory is UV complete~\cite{deform-irrelevant_py16}, not merely an EFT. In fact, this argument also holds true here, and we further show that this UV theory is free of ghosts and tachyons, hence a physical relativistic theory. Yet, the scaling dimension of the dissipation effect observed at the UV fixed point clearly deviates from what a thermodynamic \emph{flutuation-dissipation relation} (FDR) would predict. Our finding thus demonstrates a key difference between thermodynamics and driven-dissipative dynamics. 

Note that although deviations of OQS from thermodynamics have been widely studied~\cite{keldysh_sbd16} by explicit use of the Keldysh path integral, to our best knowledge, no similar studies have been done using CFT tools. Our observation implies that driven-dissipative dynamics is much richer than thermodynamics, differing in not only its non-compliance with thermodynamic symmetry~\cite{keldysh-therm_scgtd15} but its UV/IR relevance as well. In fact, although a UV-relevant dissipation effect seems unphysical in thermodynamics, it has never, in theory, been forbidden in driven-dissipative dynamics, which could remind  us Gell-Mann's totalitarian principle: ``\emph{Everything not forbidden is compulsory}''~\cite{gell-mann-total_gm56}. Our results may shed light on field-theoretical relativistic collapse models where the breakdown of unitarity may indeed happen at some nontrivial high energy scale. Our results may also offer a better understanding of universal dynamical phase transitions of condensed matters which are quasirelativistic (the dynamical exponent $z\approx1$) near the critical point. 
Finally, our results pave a new path that brings the Keldysh formalism to the gravitational side under the pronounced holographic AdS/CFT correspondence~\cite{ads-cft_m99,ads-cft_k16,ads-cft-scatter_g00,ads-cft-scatter_fk11,ads-cft-renom-group_bgl13}. This may lead to a refreshing perspective of how a Keldysh CFT living on the boundary corresponds to the bulk theory, which we will briefly discuss at the end.

\section{Keldysh formalism}
We start with a Keldysh path integral of the most general form~\cite{keldysh-i_hlr17,keldysh-ii_hlr17} for a real scalar field $\phi$,
\begin{eqnarray}
\label{eq_keldysh-path-int}
\int_{
\phi_{\pm}(t^{i}, \mathbf{x})=\phi_{\pm}^{i}(\mathbf{x})
}^{
\phi_{\pm}(t^{f}, \mathbf{x})=\phi_{\pm}^{f}(\mathbf{x})
}
D\phi_{+} D\phi_{-} e^{i S[\phi_{+},\phi_{-}]},
\end{eqnarray}
which is identified as the twofold time-evolution amplitude
\[\left\langle\phi_{+}^f(\mathbf{x})\right|\left[\mathcal{V}(t^f,t^i)(\left|\phi_{+}^i(\mathbf{x})\right\rangle\left\langle\phi_{-}^i(\mathbf{x})\right|)\right]\left|\phi_{-}^f(\mathbf{x})\right\rangle\] between the initial and final boundary conditions $\phi^{i}_{\pm}(\mathbf{x})$ and $\phi^{f}_{\pm}(\mathbf{x})$.
The time-evolution superoperator $\mathcal{V}(t^f,t^i)(\cdot)$ governs the dynamics of not bras or kets in the Hilbert space, but 
operators (matrices). The Keldysh action $S[\phi_{+},\phi_{-}]$ as a functional 
can be constructed given any specific form of $\mathcal{V}$ following the standard path integral approach (Appendix~\ref{append_s-from-l}). 

Given an initial distribution $\rho(t^i)$ as a functional of $\phi^{i}_{\pm}(\mathbf{x})$, multipoint functions of the final distribution $\rho(t^f)=\mathcal{V}(t^f,t^i)(\rho(t^i))$
are well defined by Eq.~(\ref{eq_keldysh-path-int}),
\begin{eqnarray}
\label{eq_keldysh-j}
&&\left\langle\phi_{+}(x_1)\cdots\phi_{+}(x_m)\phi_{-}(x_1)\cdots\phi_{-}(x_n)\right\rangle\nonumber\\
&=&\left.\frac{\left(-i\right)^{m+n}\delta^{m+n} Z[J_{+},J_{-}]}{\delta J_{+}(x_1)\cdots \delta J_{+}(x_m) \delta J_{-}(x_1)\cdots \delta J_{-}(x_n)}\right|_{J_{+}=J_{-}=0}
\end{eqnarray}
in terms of $\phi_{\pm}$, where the partition function
\begin{eqnarray}
\label{eq_keldysh-z}
Z[J_{+},J_{-}]=
\hspace{-5mm}
\int\limits_{\rho(t^i)}^{\phi^f_{+}(\mathbf{x})=\phi^f_{-}(\mathbf{x})}
\hspace{-5mm}
D\phi_{+} D\phi_{-} e^{i \left(S[\phi_{+},\phi_{-}]+ J_{+}\phi_{+}+ J_{-}\phi_{-}\right)}\quad
\end{eqnarray}
is constructed by adding the source terms $J_{\pm}$.
The multi-point function in Eq.~(\ref{eq_keldysh-j}) is equal to
\begin{eqnarray}
\label{eq_keldysh-time-order}
\text{Tr}\left\{\mathcal{T}\left\{\phi(x_1)\cdots\phi(x_m)\right\}\rho(t^i)\mathcal{\bar{T}}\left\{\phi(x_1)\cdots\phi(x_n)\right\}\right\}
\end{eqnarray}
in the operator formalism (Appendix~\ref{append_s-from-l}), in which $\mathcal{V}$ is invoked in $\phi(t,\mathbf{x})=\mathcal{V}^\dagger(t, t^i)(\phi(t^i,\mathbf{x}))$, given $\mathcal{V}^\dagger$ the adjoint superoperator of $\mathcal{V}$~\cite{q-open-syst}.
Note that $\rho(t^i)$ is not necessarily a density matrix. Being so, however, 
allows Eq.~(\ref{eq_keldysh-time-order}) to be interpreted as a positive and normalized physical correlation function. Also, note that two independent time-ordering and anti-time-ordering superoperators $\mathcal{T}$ and $\mathcal{\bar{T}}$ are present in Eq.~(\ref{eq_keldysh-time-order}). Thus, combinations of local operators do not necessarily follow one single time order~\cite{keldysh-renorm_bjlr17}---compared to traditional Feynman path integrals where operators can only be combined in one time order.
Indeed, Eq.~(\ref{eq_keldysh-time-order}) is the only form of correlation functions that are directly measurable~\cite{q-noise} (in contrast to the out-of-time-order correlation functions which are not~\cite{out-of-time-order-correl_hmy17}), indicating that any measurable quantum evolution can be completely described by $S[\phi_{+},\phi_{-}]$.

Here, we write down an explicit functional form for the Keldysh action, $S[\phi_{+},\phi_{-}]=\int dx L(\phi_{+},\phi_{-})$, where
\begin{eqnarray}
	\label{eq_keldysh-general}
	&&L(\phi_{+},\phi_{-})=L_0(\phi_{+})-L_0(\phi_{-})\nonumber\\
	&-&i\sum\nolimits_{i} \gamma_i \left(\mathcal{O}_{i,+}\mathcal{O}^*_{i,-}-\frac{1}{2}\mathcal{O}^*_{i,+}\mathcal{O}_{i,+}-\frac{1}{2}\mathcal{O}^*_{i,-}\mathcal{O}_{i,-}\right)\qquad
\end{eqnarray}
is locally composed of an unperturbed unitary Lagrangian $L_0(\phi)$ and \emph{arbitrary} complex scalar fields $\mathcal{O}_{i}$ as perturbations, and our metric signature convention is $\left(+,-,-,-\right)$. We require Eq.~(\ref{eq_keldysh-general}) to be Lorentz covariant; we also require that $\gamma_i\ge0$ and is time independent. As a result, the special form of Eq.~(\ref{eq_keldysh-general}) guarantees that our $S[\phi_{+},\phi_{-}]$ is at least perturbatively a field theory that obeys some Weinbergian-like constraints~\cite{weinberg-q-field-theor}, as explained below.

First, we must reverse engineer the superoperator $\mathcal{V}$ from Eq.~(\ref{eq_keldysh-general}). 
When $\mathcal{O}_{i}$ does not contain spacetime derivatives, one derives  \[\mathcal{V}(t^f,t^i)=e^{\left(t^f-t^i\right)\mathcal{L}}\] where the 
superoperator $\mathcal{L}$ 
admits a Lindblad form~\cite{q-open-syst}
\begin{eqnarray}
\label{eq_master}
&&\mathcal{L}(\rho)=-i\left[H,\rho\right]\nonumber\\
&+&\sum\nolimits_{i}\gamma_i \int d\mathbf{x}
\left(\mathcal{O}_i(\mathbf{x})\rho\mathcal{O}^\dagger_i(\mathbf{x})
-\frac{1}{2}\{\mathcal{O}^\dagger_i(\mathbf{x})\mathcal{O}_i(\mathbf{x}),\rho\}
\right),\hspace{5mm}
\end{eqnarray}
given $H$ the corresponding Hamiltonian of $L_0$ by the Legendre transformation (Appendix~\ref{append_s-from-l}). Note that in a proper Lindblad form, $\mathcal{O}_i(\mathbf{x})$ are required to be trace-zero and orthogonal to each other, but this can indeed be made so for arbitrary $\mathcal{O}_i(\mathbf{x})$ by linear recombination~\cite{lindblad-derive_p12}. This explains why there is no constraint on $\mathcal{O}_i$ in Eq.~(\ref{eq_keldysh-general}).

When $\mathcal{O}_{i}$ contains spacetime derivatives (which is not forbidden since unitarity is not concerned), $\mathcal{V}$ is intractable and may not be Lindblad as wished. Nevertheless, \emph{Matthews's theorem}~\cite{matthews-theor_m49} implies that the na\"ive Lorentz-covariant Lagrangian path-integral formalism [Eq.~(\ref{eq_keldysh-general})] should yield at least perturbatively identical results compared to the correct Hamiltonian path-integral formalism [Eq.~(\ref{eq_master})] even if $\mathcal{O}_i$ contains spacetime derivatives~\cite{matthews-theor-first-order_bd75,matthews-theor-high-order_k94}. Thus, Eq.~(\ref{eq_master}) is still valid perturbatively. 

Note that our $\mathcal{V}$ is Markovian, i.e., $\mathcal{V}(t_1+t_2,0)=\mathcal{V}(t_1,0)\mathcal{V}(t_2,0)$, for $t_1,t_2>0$, so that the evolution of the system in the future does not depend on its history, as demanded by locality and translational invariance~\cite{q-open-syst}. Therefore, $\{\mathcal{V}(t,0)\}$ as a one-parameter family of $t$ forms a quantum dynamical semigroup. One may suspect if the semigroup can still produce some sort of conservation law. The answer is yes~\cite{q-dyn-semigroup-noether_grs15}. 
Equation~(\ref{eq_master}) has been widely used to describe the dynamics of an OQS under the Born-Markov approximation, which means that rich physics of nonlocality (non-Markovianity~\cite{nMarkov-colloq_blpv16}) such as decrease of quantum speed limit~\cite{q-speed-limit_mwg15} or emergence of multiple timescales~\cite{multi-scale_mlzgs19} has to be ignored. In the language of quantum measurement, this amounts to the assumption that the nonselective continuous measurements the OQS undergoes are weak enough~\cite{q-open-syst}. It is, however, unclear to what extent non-Markovianity should be seriously considered in a fundamental field theory of some kind, e.g., quantum collapse models~\cite{q-collapse_blssu13}, where a violation of locality is perhaps unfavorable.

We stress that our $\mathcal{V}$ is known to obey the following two ``\emph{CPT}'' constraints:

\subsection{CPT (complete positivity and trace persevering)}
Any physical $\mathcal{V}$ that maps density matrices to density matrices must be complete positive and trace persevering~\cite{q-open-syst}. The Choi-Kraus's theorem~\cite{ck-theor_c75,fundam-notion-q-theor} states that an arbitrary superoperator $\mathcal{V}(\rho)$ is 
\emph{CPT} 
if and only if it can be expressed as 
$\mathcal{V}(\rho)=\sum\nolimits_{i}\Omega_{i}\rho \Omega_{i}^\dagger$ where $\Omega_{i}$ is a bounded Kraus operator that satisfies $\sum\nolimits_{i}\Omega_{i}^\dagger\Omega_{i}=1$.
The Gorini-Kossakowski-Sudarshan-Lindblad theorem~\cite{gksl-theor_gks76,gksl-theor_l76} then states that if $\{\mathcal{V}(t^f,t^i)\}$ forms a dynamical semigroup, then it satisfies the Choi-Kraus's form if and only if it can be written as $e^{\mathcal{L}\left(t^f-t^i\right)}$ with a Lindblad form $\mathcal{L}$---which is the same as in Eq.~(\ref{eq_master}).
A caveat, however, lies in the fact that operators in field theories are almost always unbounded, and hence, they have to be properly regularized before the Choi-Kraus's theorem being applied.

\subsection{CPT (charge, parity, and time reversal symmetry)}
A \emph{CPT} transformation $\Theta$ may be defined as any one of the various antiunitary transformations of a theory that has some global invariance (e.g., a Poincar\'e invariance)~\cite{keldysh-i_hlr17}. In fact, from the Choi-Kraus's theorem immediately comes an invariance under (antiunitary) complex conjugation, i.e., $\text{Tr}\{\mathcal{V}(\rho)\}^*=\text{Tr}\{\sum\nolimits_{i}\Omega_{i}\rho^\dagger \Omega_{i}^\dagger\}=\text{Tr}\{\sum\nolimits_{i}\Omega_{i}\rho\Omega_{i}^\dagger\}=\text{Tr}\{\mathcal{V}(\rho)\}$. 
There are different ways to realize $\Theta$ in the path-integral formalism: for example, when $\mathcal{O}^*_{i}=\mathcal{O}_{i}$, Eq.~(\ref{eq_keldysh-general}) explicitly obeys a $C$ symmetry by remaining invariant under $L(\phi_{+},\phi_{-})\to L'(\phi_{+},\phi_{-})=L(\phi_{+}^*,\phi_{-}^*)$, and thus, $\Theta$ can be realized as $Z[J_{+},J_{-}]\to Z'[J_{+},J_{-}]= Z^*[-J_{-}^*,-J_{+}^*]$, the invariance of which becomes a \emph{CPT} symmetry~\cite{keldysh-i_hlr17}. Note that under this realization, $\Theta$ does not actually reverse the time label $t\to-t$ but rather just swap the bra-ket subscripts, because $\{\mathcal{V}(t,0)\}$ cannot have a time-reversal symmetry since it is only a semigroup, not a group.

\section{Double-trace deformation}

In the following, w.l.o.g.,~we will restrict ourselves to only one $\gamma$ term. We will also demand 
$\mathcal{O}^*=\mathcal{O}$ and save the discussion of $C$-symmetry-breaking terms for later. Instead of the bra-ket basis, a so-called Keldysh basis is particularly useful by defining 
\begin{eqnarray*}
\mathcal{O}_c&=&\left(\mathcal{O}_{+}+\mathcal{O}_{-}\right)/\sqrt{2},\text{ and }\nonumber\\ \mathcal{O}_q&=&\left(\mathcal{O}_{+}-\mathcal{O}_{-}\right)/\sqrt{2}
\end{eqnarray*}
for arbitrary fields $\mathcal{O}_{\pm}$. The subscripts $c$ and $q$ correspond to ``classical'' and ``quantum'', respectively, as they are individually and closely related to the classical and quantum contributions to the noise spectrum in the theory of quantum noise~\cite{q-meas_cdgms10}. We will work directly in this basis for matrix representations throughout the context. 

After rotated into the Keldysh basis, the full Lagrangian Eq.~(\ref{eq_keldysh-general}) reads
\begin{eqnarray}
\label{eq_keldysh-general-cq}
&&L(\phi_{+},\phi_{-})=L_0(\phi_{+})-L_0(\phi_{-})+i \gamma \mathcal{O}_{q}^2.
\end{eqnarray}
We let the coupling parameter scale as $\gamma\sim\Lambda^{d-2\Delta}$ assuming that $\mathcal{O}_{q}$ is a single-trace operator of scaling dimension $\Delta$, thus identifying $\Lambda$ as the cutoff of the theory. 
As a result, one immediately recognizes that Eq.~(\ref{eq_keldysh-general-cq}) is nothing but a double-trace deformation~\cite{deform-ads-cft_w02,deform_gk03,deform_a10,deform_gkp18} by $\mathcal{O}_{q}^2$ on the undeformed part $L_0(\phi_{+})-L_0(\phi_{-})$. The only subtlety is that the double-trace deformation in Eq.~(\ref{eq_keldysh-general-cq}) must be considered as not a single scalar but a multiplet, the coupling strength of which is a two-by-two matrix
$\label{eq_g22_f}
{\mathbf{f}}=\begin{pmatrix}
0 & 0\\
0 & -2i \gamma
\end{pmatrix}
$ 
in the Keldysh basis. Such a multiplet deformation gives rise to a recombination of different operators (in our case, $\mathcal{O}_c$ and $\mathcal{O}_q$) that occurs along the RG flow by tuning $\gamma$~\cite{deform-multiplet_bbdpr16}. This can be most effectively understood in the large $N$ limit where all multipoint connected functions are automatically suppressed by powers of $N$~\cite{ads-cft_m99}. Only left are the two-point functions. 

\subsection{Two-point functions in general}
Recall that given a general $S[\phi_{+},\phi_{-}]$, the two-point functions
\[\left\langle{\mathbf{O}}(x){\mathbf{O}}^\intercal(0)\right\rangle=
\int \frac{d^d k}{\left(2\pi\right)^d}
\mathbf{G}(k)
e^{-ikx}\] for an arbitrary $\mathcal{O}$ [with shorthand ${\mathbf{O}}=\left(\mathcal{O}_{c},\mathcal{O}_{q}\right)^\intercal$]
encode all possible second-order correlations in the form of Eq.~(\ref{eq_keldysh-time-order}), provided that $\rho(t^i)$ in Eq.~(\ref{eq_keldysh-time-order}) is identified as a stationary state $\rho_0$ of $S[\phi_{+},\phi_{-}]$~\cite{keldysh_sbd16}. 
For a unitary QFT,  with the infinitesimal Wick rotation understood, $\rho_0$ is just the vacuum state. For nonunitary dynamics, $\rho_0$ is not necessarily a thermal equilibrium state.
In the momentum space, one has
\begin{eqnarray*}
\mathbf{G}(k)=
\begin{pmatrix}
	G^{K}(k) & G^{R}(k)\\
	G^{A}(k)&0
\end{pmatrix}.
\end{eqnarray*}
There is a redundancy reflected by the $\mathcal{O}_q$-$\mathcal{O}_q$ correlation being nilpotent~\cite{keldysh_sbd16}. In fact, any correlation function composed of only $\mathcal{O}_q$ operators will be nilpotent, a reflection of the conservation law of probability~\cite{keldysh-i_hlr17,keldysh-ii_hlr17}.
Noticing the use of the Keldysh basis and the cyclic property of trace in Eq.~(\ref{eq_keldysh-time-order}), one can proceed to see that $G^{R/A}$ are exactly the \emph{retarded/advanced functions} that by definition describe the linear response of the system. The \emph{Keldysh function} $G^K$, on the other hand, describes the spectrum of the symmetric auto-correlation of $\phi$~\cite{keldysh_sbd16}.

Would the underlying dynamics of a general Keldysh theory be \emph{thermal}, there would be a universal relation between $G^{R/A}$ and $G^K$, namely, the FDR. In that case, $\left\langle{\mathbf{O}}(x){\mathbf{O}}^\intercal(0)\right\rangle$ could be derived alternatively by the Matsubara formalism~\cite{therm-field-theor}: In short, one introduces a Euclidean action $S^E[\phi;\beta]=-\int_{0}^{\beta}d\tau\int d\mathbf{x}L_0(\phi)$ with periodicity $\beta$ (inverse temperature) in the imaginary time direction $\tau=it$. From $S^E[\phi;\beta]$ a Euclidean Green's function $G^E(\omega_n, \mathbf{k})$ defined at discrete Matsubara frequencies $\omega_n=2 \pi n/\beta$, $n\in\mathbb{Z}$ can be derived, which is connected to the Minkowski-space $G^{R/A}$ through analytical continuation,
\begin{eqnarray}
\label{eq_matsubara}
G^{R}(\omega, \mathbf{k})&=&-i G^{E}(-i\omega+\varepsilon, \mathbf{k}),\nonumber\\
G^{A}(\omega, \mathbf{k})&=&-i G^{E}(-i\omega-\varepsilon, \mathbf{k}). 
\end{eqnarray}
Then, using the Kubo-Martin-Schwinger condition~\cite{keldysh-therm_scgtd15}, a connection between the fluctuation spectrum $G^K$ and the susceptibility $G^R-G^A$ is given by
\begin{eqnarray}
\label{eq_fdr}
G^{K}&=&\coth\left(\beta\omega/2\right)\left(G^{R}-G^{A}\right)\text{ (only if thermal)    }
\end{eqnarray}
which is nothing but the FDR~\cite{keldysh_sbd16}. When $\beta\to\infty$, Eq.~(\ref{eq_fdr}) reduces to $G^{K}=\text{sign}(\omega)\left(G^{R}-G^{A}\right)$ which always holds for unitary QFTs.

\subsection{RG flow induced by deformation}

Now we calculate $\mathbf{G}$ for our deformed Lagrangian [Eq.~(\ref{eq_keldysh-general-cq})]. Following the large-$N$ approach, we find, in the large $N$ limit (Appendix~\ref{append_deform}),
\begin{eqnarray}
\label{eq_g22}
\mathbf{G}\simeq\left(\mathbf{G}_0^{-1}+i\mathbf{f}\right)^{-1}
\simeq\begin{pmatrix}
G^{K}_{0}-2\gamma G^{R}_{0}G^{A}_{0} & G^{R}_{0}\\
G^{A}_{0} & 0
\end{pmatrix}
\end{eqnarray}
where $G^{R/A}_{0}$ and $G^{K}_{0}$ are components of $\mathbf{G}_0$ for the undeformed part $L_0(\phi_{+})-L_0(\phi_{-})$. We find that, in the large $N$ limit, $G^{R/A}=G^{R/A}_{0}$ remains unchanged, suggesting that the deformation term $i\gamma\mathcal{O}_q^2$ does not modify the susceptibility of the system. The new Keldysh function, 
$G^{K}=G^{K}_{0}+G^{K}_{\mathcal{L}}$,
however, contains a deformation-induced correction which we define as $G^{K}_{\mathcal{L}}=-2\gamma G^{R}_{0}G^{A}_{0}$. 

If we tune the cutoff $\Lambda$ from zero to infinity, then it not only changes the dimensional coupling $\gamma$ but also drives the theory to flow between the two ends of the RG trajectory, which we assume to be two different CFTs.
If we let $L_0$ be a unitary CFT, then $G^{R/A}_{0}$ and $G^{K}_{0}$ for the vacuum (i.e., $\beta=\infty$) can be derived from Eqs.~(\ref{eq_matsubara})~and~(\ref{eq_fdr}),
\begin{eqnarray}
\label{eq_cft-g}
&G^{R/A}_0&=-iC\left[\mathbf{k}^2-\left(\omega\pm i\varepsilon\right)^2\right]^{\Delta-d/2},\text{ and}\nonumber\\
&G^{K}_0&=\begin{cases} 
2C\sin\left[(d/2-\Delta)\pi\right]\left(k^2\right)^{\Delta-d/2} & 
k^2>0\nonumber\\
0 & 
k^2<0
\end{cases},
\end{eqnarray}
with $C=4^{d/2-\Delta}\pi^{d/2}\Gamma(d/2-\Delta)/\Gamma(\Delta)$~\cite{deform-irrelevant_py16} given $k^2=\omega^2-\mathbf{k}^2$. Condition $\Delta-d/2\geq-1$ should be imposed as required by unitarity~\cite{conform-field-theor}. Special care should also be taken when $\Delta-d/2\in\mathbb{Z}$ since the results above will have log corrections.
Apparently, $G^{R/A}_0\sim G^{K}_0\sim \left|k\right|^{2\Delta-d}$ should have the same scaling behavior, as they come from the same unitary operator $\mathcal{O}$. Note that the CFT $L_0(\phi_{+})-L_0(\phi_{-})$ is not just a sum of two identical and independent Minkowski CFTs: twisted fields (since there is an internal symmetry between $\phi_{\pm}$) that connect the two CFTs through branch points should also exist~\cite{twist-field_ccad08}, which should be responsible for the nonzero $G^{K}_0$.

The RG flow is triggered after $\gamma$ is turned on. We find
\[G^{K}_{\mathcal{L}}=2\gamma C^2 \left|k^2\right|^{2\left(\Delta-d/2\right)},\] which scales with $\left|k\right|$ but differently than $G^{K}_0$ does. Hence, $G^{K}=G^{K}_{0}+G^{K}_{\mathcal{L}}$ cannot obey the FDR [Eq.~(\ref{eq_fdr})]. This is the main difference between a thermal field theory and a driven-dissipative field theory: the former always obeys the FDR as guaranteed by microscopic unitarity, yet the latter describes driven-dissipative dynamics which usually exhibits nonthermodynamic characteristics~\cite{keldysh_sbd16}. We thus conclude that $\gamma\to\infty$ leads to a new, nonunitary, and nonthermal CFT. The new scaling dimension for $\mathcal{O}_c$ can be identified as $\Delta_c|_{\gamma\to\infty}=2\Delta-d/2$.

Now we emphasize the key observation: at the $O(N^0)$ level, the Keldysh function $G^{K}$ remains positive (free of ghosts) and finite (free of tachyons) and hence, \emph{physical}~\cite{deform-irrelevant_py16} if and only if $\gamma>0$, a condition which nevertheless has been guaranteed by our general construction of the Keldysh theory. Therefore, Eq.~(\ref{eq_g22}) defines a complete RG trajectory---along which the theory works at \emph{all} energy scales and may flow in both way between UV and IR fixed points~\cite{deform-irrelevant_py16}. In particular, when $\Delta>d/2$, Eq.~(\ref{eq_g22}) yields a UV fixed point at $|k|/\Lambda\to\infty$ which the IR-irrelevant deformation $\sim{\Lambda^{d-2\Delta}}\mathcal{O}_q^2$ will eventually flow backwards to. This is to say that our irrelevantly perturbed Keldysh theory can be rephrased into a new nonunitary field theory which is a relevant perturbation of the UV fixed point. Such UV \emph{completion} has been well known and studied for unitary QFTs, e.g., $O(N)$ vector model near $d=6$~\cite{phi4-uv-complete_fgk14}, yet now we have seen that a Lindblad-form deformation could also be UV completed in the same manner.
This observation is unexpected, since a dissipation-like perturbation which breaks unitarity
was always expected to be IR relevant, and 
thus, \emph{irreversibility} should only be observed at the macroscopic scale~\cite{keldysh-therm_scgtd15}. This IR-relevant behavior corresponds to $\Delta<d/2$ in our driven-dissipative field theory [Eq.~(\ref{eq_keldysh-general-cq})]. Yet as we have seen, $\Delta>d/2$ neither contradicts the Weinbergian requirements of the Keldysh theory nor denies its renormalizability. Driven-dissipative quantum dynamics is hence much {richer} than thermodynamics, differing not only in whether complying with the FDR but in the UV/IR relevance as well.

\begin{figure}[t]

	\begin{minipage}[t]{36mm}
		{\includegraphics[height=28mm]{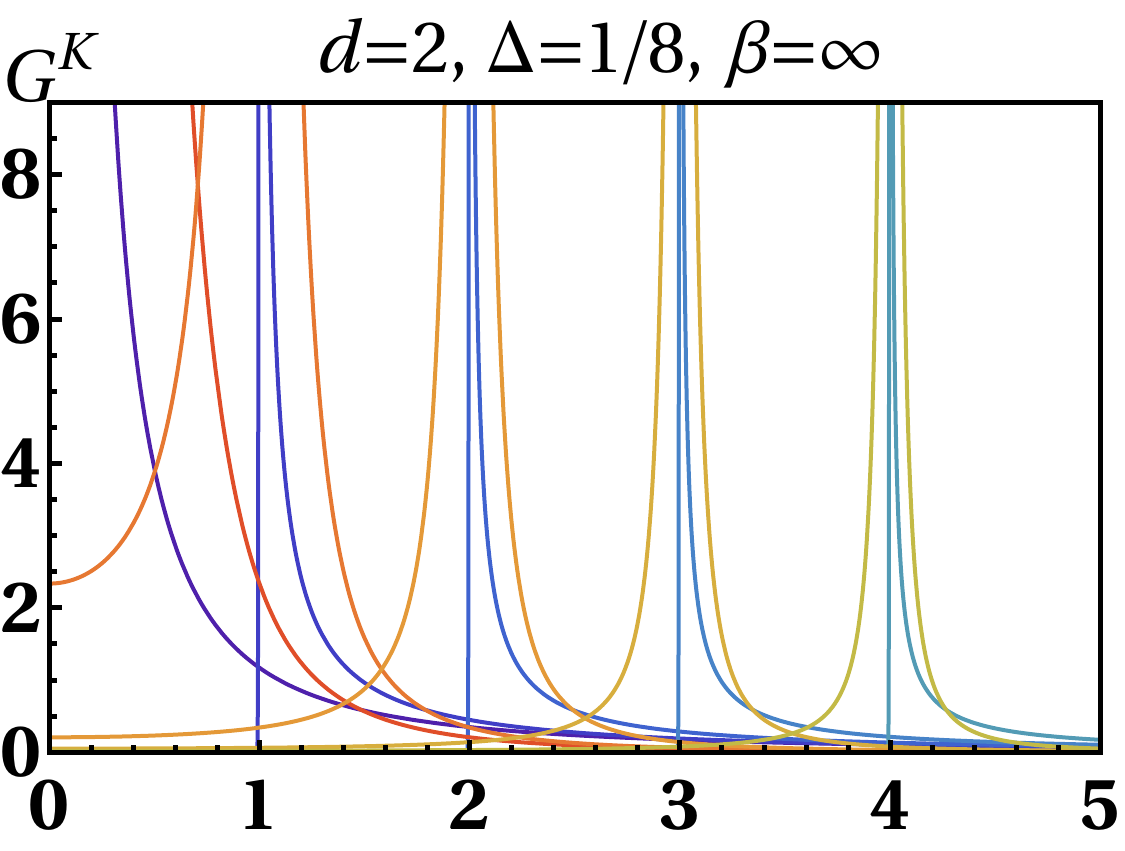}
			\vspace{-6mm}
			\subcaption{\label{fig_beta-inf}}}
	\end{minipage}
	\begin{minipage}[t]{47mm}
		{\includegraphics[height=28mm]{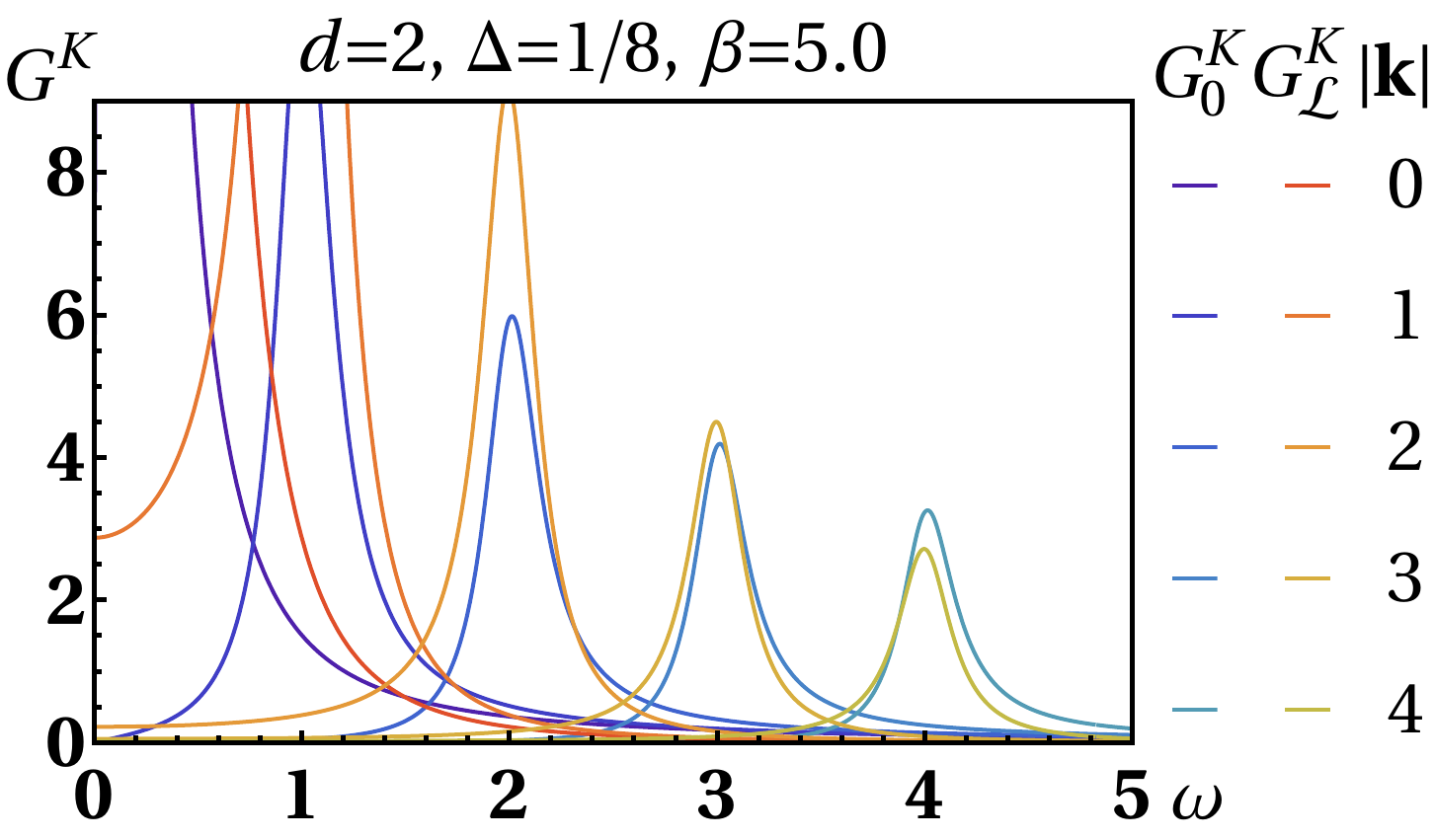}
			\vspace{-6mm}
			\subcaption{\label{fig_beta-5}}}
	\end{minipage}
	
	\caption{\label{fig_beta}The Keldysh function $G^{K}_0+G^{K}_{\mathcal{L}}$ consists of two parts: $G^{K}_0$ is the undeformed Keldysh function; $G^{K}_{\mathcal{L}}$ is the correction induced by double-trace deformation $i\gamma\mathcal{O}^2$. Both $G^{K}_0$ and $G^{K}_{\mathcal{L}}$ are even functions of $\omega$ and are exactly solvable in the large $N$ limit  in two dimensions  for arbitrary temperature $\beta$: \subref{fig_beta-inf}~$\beta=\infty$; \subref{fig_beta-5}~$\beta=5.0$. Both have $\gamma=0.5$. The scaling dimension of $\mathcal{O}$ is $\Delta=1/8$.\hfill\hfill}
\end{figure}

The large-$N$ approach is also applicable to thermal CFTs. For $d=2$, multipoint functions at finite temperature $\beta^{-1}$ can be calculated from the vacuum by the Weyl transformation~\cite{weyl-vs-conform_flp17}. Putting exactly known results~\cite{therm-q-ising_sss94,therm-q-ising_s97} into Eqs.~(\ref{eq_matsubara})~and~(\ref{eq_fdr}), we derive $G^{K}_0=2{\Im}\left\{f(k,\beta)\right\}$ and  $G^{K}_{\mathcal{L}}=2\gamma \left|f(k,\beta)\right|^2$,
where
\begin{eqnarray*}
&& f(k,\beta)=\\
&& -C \left(\frac{\beta}{4\pi}\right)^{2-2\Delta}
\frac{\Gamma(\frac{\Delta}{2}+i\frac{\beta\left(\omega+\left|\mathbf{k}\right|\right)}{4\pi})\Gamma(\frac{\Delta}{2}+i\frac{\beta\left(\omega-\left|\mathbf{k}\right|\right)}{4\pi})}{\Gamma(\frac{2-\Delta}{2}+i\frac{\beta\left(\omega+\left|\mathbf{k}\right|\right)}{4\pi})\Gamma(\frac{2-\Delta}{2}+i\frac{\beta\left(\omega-\left|\mathbf{k}\right|\right)}{4\pi})}.
\end{eqnarray*}
We compare $G^{K}_0$ and $G^{K}_{\mathcal{L}}$ for $\beta=\infty$ and $\beta=5.0$ in Fig.~\ref{fig_beta}, choosing $\Delta=1/8$ (which is the scaling dimension of the $\mathbb{Z}_2$-odd field $\sigma$ in $d=2$ Ising CFT) and $\gamma=0.5$. We clearly see that $G^{K}_0$ and $G^{K}_{\mathcal{L}}$ have different scaling behaviors and thus $G^{K}_{\mathcal{L}}$ cannot obey the FDR.

We finally discuss in brief what if $\mathcal{O}$ is charged: a $C$-symmetry-breaking term \[i\gamma\left(\mathcal{O}_c\mathcal{O}^*_q-\mathcal{O}^*_c\mathcal{O}_q\right)/2\] will appear in Eq.~(\ref{eq_keldysh-general-cq}), which will alter $G^{R/A}$ as well, 
by shifting the poles in $G^{R/A}_0$ to both upper and lower half planes, causing the retarded/advanced functions $G^{R/A}$ to become \emph{unphysical}. This is because in a relativistic scalar field theory, particles and antiparticles always come in pairs so that the charge spectrum is not bounded below---and thus a dissipation of charge will blow up in either time direction, which is a crucial difference from previously studied nonrelativistic field theories~\cite{keldysh_sbd16}. 

\section{Examples}

\subsection{$\left(0+1\right)$-dimensional massive scalar field}

Our first example is simply quantum mechanical: a one-dimensional quantum harmonic oscillator of intrinsic frequency $m$ under weak, continuous nonselective position $q$ measurement with rate $\gamma$~\cite{q-open-syst}. This is the simplest model that does not admit a quantum nondemolition measurement~\cite{q-meas_cdgms10}, since the back action of measurement will affect the correlation functions of the observable $q$.
The Lagrangian deformed by $\mathcal{O}_q^2=\left(q_{+}-q_{-}\right)^2/2$ is given by
\begin{eqnarray}
\label{eq_1dho-lagrangian}
L(q_{+},q_{-})&=&\frac{1}{2}\left[\left(\partial_t q_{+}\right)^2-m^2 q_{+}^2\right]-\frac{1}{2}\left[\left(\partial_t q_{-}\right)^2-m^2 q_{-}^2\right]\nonumber\\
&&-i\gamma\left(q_{+}q_{-}-\frac{1}{2}q_{+}q_{+}-\frac{1}{2}q_{-}q_{-}\right),
\end{eqnarray}
which is quadratic. Thus, Eq.~(\ref{eq_g22}) is exact even when $N=1$ as in our example, and therefore,
$G^{R/A}=i/[(\omega\pm i\varepsilon)^2-m^2]$
remains unchanged under the deformation. This can also be seen directly by a Gaussian integral of Eq.~(\ref{eq_1dho-lagrangian}). Meanwhile, we have $G^{K}_{\mathcal{L}}=-2\gamma G^{R}G^{A}\sim \omega^{-4}$ by dimensional analysis.
A more careful calculation further yields an approximate relation (Appendix~\ref{append_qho}),
\begin{eqnarray*}
	G^{K}_{\mathcal{L}}\simeq\frac{\gamma}{m\varepsilon}\pi\delta\left(\omega^2-m^2\right)=\frac{\gamma}{2\omega\varepsilon}\left(G^{R}-G^{A}\right). 
\end{eqnarray*}
This is not an exact FDR since we are not looking at a thermal theory, but it looks very similar to Eq.~(\ref{eq_fdr}). Hence, we can introduce a ``sloppy'' effective inverse temperature $\beta_\text{eff}\simeq 4\varepsilon/\gamma-\left(8m \varepsilon^2/\gamma^2\right) \coth \left(\beta m/2\right) +O(\varepsilon^3) $ regulated by $\varepsilon$ and rewrite the Keldysh two-point function as
$G^{K}\simeq
\left[\coth\left(\beta \omega/2\right)+{\gamma}/{\left(2 \omega\varepsilon\right)}\right]
\left(G^{R}-G^{A}\right)\simeq\coth\left(\beta_\text{eff} \omega/2\right)\left(G^{R}-G^{A}\right)$. 
Thus, $G^{K}$ effectively describes the fluctuation of an infinitely heated harmonic oscillator, which is a stationary state of Eq.~(\ref{eq_1dho-lagrangian}), in consistence with the Hamiltonian dynamics (Appendix~\ref{append_qho}). This steadily infinite heating is also a well-known feature of collapse models~\cite{q-collapse_blssu13}. Interestingly, we see that in the Keldysh formalism, the role of the regulator is played by $\varepsilon$ instead of the spatial localization width $r_C$ commonly used in those models~\cite{q-collapse_blssu13}.

\subsection{$\left(4-\epsilon\right)$-dimensional $O(N)$ massless scalar field}
The second example we are interested in is described by
\begin{eqnarray}
\label{eq_nscalar-lagrangian}
&& L(\phi^{\alpha}_{+},\phi^{\alpha}_{-})=L_0(\phi^{\alpha}_{+})-L_0(\phi^{\alpha}_{-})\nonumber\\
&&-i\gamma\left[
\left(\phi^{\alpha}\phi^{\alpha}\right)_{+}\left(\phi^{\alpha}\phi^{\alpha}\right)_{-}-\frac{1}{2}\left(\phi^{\alpha}\phi^{\alpha}\right)_{+}^2-\frac{1}{2}\left(\phi^{\alpha}\phi^{\alpha}\right)_{-}^2\right],\qquad
\end{eqnarray}
where $L_0(\phi^{\alpha})=\frac{1}{2}\partial_{\mu} \phi^{\alpha}\partial^{\mu} \phi^{\alpha}
-\frac{\lambda}{4!}\left(\phi^{\alpha}\phi^{\alpha}\right)^2$ is an ordinary $O(N)$ massless scalar field theory. 
In the Keldysh basis, we have
$2\mathcal{O}_q^2=\left(\mathopen{:}\phi^{\alpha}_{+}\phi^{\alpha}_{+}-\phi^{\alpha}_{-}\phi^{\alpha}_{-}\mathclose{:}\right)^2=4\left(\mathopen{:}\phi^{\alpha}_c \phi^{\alpha}_q\mathclose{:}\right)^2$ and $2\mathcal{O}_c^2=\left(\mathopen{:}\phi^{\alpha}_{+}\phi^{\alpha}_{+}+\phi^{\alpha}_{-}\phi^{\alpha}_{-}\mathclose{:}\right)^2=\left(\mathopen{:}\phi^{\alpha}_c \phi^{\alpha}_c\mathclose{:}\right)^2$. Note that $\mathcal{O}$ is normal ordered for being a single-trace operator, which should always be understood below. If $\gamma$ remains zero, then the real and positive coupling $\lambda$ will trigger an RG flow from the Gaussian fixed point at $\lambda=0$ to the Wilson-Fisher (WF) fixed point at $\lambda/4!\simeq\mu^{\epsilon}{2\pi^2 \epsilon}/\left({N+8}\right)+O(\epsilon^2)$ by expansion around $\epsilon=4-d$ 
at an energy scale $\mu$~\cite{stat-field-theor-1,stat-field-theor-2}. 
Note that when $\epsilon> 0$, the stationary states at both fixed points are stable.

Now we turn on $\gamma$ in Eq.~(\ref{eq_nscalar-lagrangian}). Diagrammatic perturbative calculations can be carried out in the same spirit of the usual counterterm renormalization scheme. Yet, the diagram counting is more complex, since we have three free propagators $\boldsymbol{\mathcal{P}}(k)$
in total in the Keldysh basis 
(Appendix~\ref{append_diagram}). 
Such an analysis has been done for $N=1$~\cite{keldysh-renorm_bjlr17}. Here, we extend it to $O(N)$  and find that the beta functions for the dimensionless renormalized parameters $\bar{\lambda}=\mu^{-\epsilon}\lambda$ and $\bar{\gamma}=\mu^{-\epsilon}\gamma$ are given by (Appendix~\ref{append_diagram})
\begin{eqnarray*}
\label{eq_nscalar-beta1}
\beta_{\bar{\lambda}}&\simeq&-\epsilon\bar{\lambda}+\frac{1}{\left(4\pi\right)^2}\frac{N+8}{3}\bar{\lambda}^2,\nonumber\\
\beta_{\bar{\gamma}}&\simeq&-\epsilon\bar{\gamma}+\frac{1}{\left(4\pi\right)^2}\frac{2N+13}{3}\bar{\lambda}\bar{\gamma},
\end{eqnarray*}
which have \emph{four} fixed points in total: there are two fixed values for $\bar{\lambda}$, which are $\bar{\lambda}_*=0$ and $\bar{\lambda}_*/4!\simeq{2\pi^2 \epsilon}/\left({N+8}\right)$; and two for $\bar{\gamma}$, which are $\bar{\gamma}_*=0$ and $\bar{\gamma}_*=\infty$. We see that both the Gaussian and the WF fixed points for $\bar{\lambda}$ for the unitary theory ($\bar{\gamma}_*=0$) are recovered. Meanwhile, $\beta_{\bar{\lambda}}$ and $\beta_{\bar{\gamma}}$ can reduce to and match the $N=1$ result~\cite{keldysh-renorm_bjlr17}. 
At the Gaussian fixed point, the scaling dimensions corresponding to the couplings $\lambda$ and $\gamma$ are given by
$\Delta^\text{Gauss}_{\phi^{\alpha}_c\phi^{\alpha}_c\phi^{\beta}_c\phi^{\beta}_q}=\Delta^\text{Gauss}_{\left(\phi^{\alpha}_c\phi^{\alpha}_q\right)^2}=4-2\epsilon$, indicating that $\mathcal{O}_c\mathcal{O}_q$ and $\mathcal{O}^2_q$ as relevant operators when $\epsilon> 0$ have the same trivial engineering dimension. At the WF fixed point, however, we find $\Delta^\text{WF}_{\phi_c^{\alpha}\phi_c^{\alpha}\phi_c^{\beta}\phi_q^{\beta}}\simeq4+O(\epsilon^2)$ and 
$\Delta^\text{WF}_{\left(\phi_c^{\alpha}\phi_q^{\alpha}\right)^2}\simeq4-{3\epsilon}/{\left(N+8\right)}+O(\epsilon^2)$. Both operators are thus irrelevant when $\epsilon> 0$ and independent.

The operators of which the scaling dimensions we are mostly interested in are $\mathcal{O}_c$ and $\mathcal{O}_q$. To this end, we add quadratic terms $-m^2\phi^{\alpha}_c\phi^{\alpha}_q$ and $ic\phi^{\alpha}_c\phi^{\alpha}_c/2$ to the free Lagrangian $L_0(\phi^{\alpha}_{+})-L_0(\phi^{\alpha}_{-})$. The first term generates a positive mass for the unitary theory. 
The second term is a pure vertex of some power of $\phi_c$---the kind of which is known to be responsible for breaking the Lindblad constraint~\cite{keldysh-renorm_bjlr17}---hence, only used for bookkeeping and should be taken to zero at the final step of calculation. Including the second term adds to the free propagator $\boldsymbol{\mathcal{P}}$ a new term $c\boldsymbol{\mathcal{P}}^{(1)}+O(c^2)$ where 
$\boldsymbol{\mathcal{P}}^{(1)}=
-\begin{pmatrix}
	\left(\mathcal{P}^{K}\right)^2 & \mathcal{P}^{R}\mathcal{P}^{K}\\
	\mathcal{P}^{A}\mathcal{P}^{K}&\mathcal{P}^{A}\mathcal{P}^{R}
\end{pmatrix}
$
is the first-order correction.
From the modified propagators we derive two identical beta functions (Appendix~\ref{append_diagram}),
\begin{eqnarray*}
\label{eq_nscalar-beta2}
\beta_{\bar{m}^2}&\simeq&-2\bar{m}^2+\frac{1}{\left(4\pi\right)^2}\frac{N+2}{3}\bar{\lambda}\bar{m}^2,\nonumber\\
\beta_{\bar{c}}&\simeq&-2\bar{c}+\frac{1}{\left(4\pi\right)^2}\frac{N+2}{3}\bar{\lambda}\bar{c},
\end{eqnarray*}
for $\bar{m}^2=\mu^{-2}m^2$ and $\bar{c}=\mu^{-2}c$, respectively. Again, we recover the beta function of mass renormalization. Note that both scaling dimensions are equal, given by
$\Delta^\text{Gauss}_{\phi^{\alpha}_c\phi^{\alpha}_q}=\Delta^\text{Gauss}_{\phi^{\alpha}_c\phi^{\alpha}_c}=2-\epsilon$ and $\Delta^\text{WF}_{\phi^{\alpha}_c\phi^{\alpha}_q}=\Delta^\text{WF}_{\phi^{\alpha}_c\phi^{\alpha}_c}\simeq2-6\epsilon/\left({N+8}\right)+O(\epsilon^2)$. This makes sense since $\mathcal{O}_c$ and $\mathcal{O}_q$ must have the same scaling dimension: they are only linear combinations of bra and ket fields, which are completely equivalent at $\bar{\gamma}_*=0$. 

Now we can calculate the retarded/advanced functions, i.e.,~the two-point functions between $\mathcal{O}_c$ and $\mathcal{O}_q$. When $N\to\infty$, we find
\begin{eqnarray}
\label{eq_nscalar-g}
& G^{R/A}_\text{Gauss}&=\left(2N\right) i C_{\epsilon}^{-1}\left[\mathbf{k}^2-\left(\omega\pm i\varepsilon\right)^2\right]^{-\epsilon/2},\nonumber\\
& G^{R/A}_\text{WF}&=-\left(2N\right)^{-1} i C_{\epsilon}\left[\mathbf{k}^2-\left(\omega\pm i\varepsilon\right)^2\right]^{\epsilon/2},
\end{eqnarray}
at the Gaussian and the WF fixed points, respectively, with
$C_{\epsilon}$ a normalization coefficient (Appendix~\ref{append_diagram}). 
This result is derived by noticing that the RG flow connecting the two unitary fixed points is also triggered by a double-trace deformation by $\lambda$~\cite{phi4-uv-complete_fgk14}, and hence, $G^{R/A}_\text{WF}$ can be easily calculated from $G^{R/A}_\text{Gauss}$  in the large $N$ limit (following Appendix~\ref{append_deform}). Note that we have dropped a contact term $\sim 1/\lambda \sim \mu^{-\epsilon}$ from $G^{R/A}_\text{WF}$ and then renormalized it by a multiplicative factor $\lambda^2 \sim \mu^{2\epsilon}$ in Eq.~(\ref{eq_nscalar-g}).

We move our focus to the nonunitary fixed points at $\bar{\gamma}_*=\infty$. 
The crucial difference between the RG flows originating from the two unitary fixed points is that the $i\gamma\mathcal{O}^2_q$ deformation is \emph{relevant} at the Gaussian fixed point but \emph{irrelevant} at the WF fixed point when $\epsilon>0$.
This can be clearly seen from Eq.~(\ref{eq_nscalar-g}).
When approaching the nonunitary fixed points, we have
\begin{eqnarray*}
 G^{K}_\text{Gauss}\xrightarrow{\gamma\to\infty}&  -2\gamma G^{R}_\text{Gauss}G^{A}_\text{Gauss} &\simeq 
2\gamma \left(2N\right)^2 C_{\epsilon}^{-2}\left|k^2\right|^{-\epsilon},\nonumber\\
G^{K}_\text{WF}\xrightarrow{\gamma\to\infty}& -2\gamma G^{R}_\text{WF}G^{A}_\text{WF} &\simeq  
2\gamma \left(2N\right)^{-2} C_{\epsilon}^2 \left|k^2\right|^{\epsilon},
\end{eqnarray*}
in the large $N$ limit. We can write down the new scaling dimensions for $\mathcal{O}_c$ at the two nonunitary fixed points:
\begin{eqnarray*}
\Delta^\text{Gauss}_{\phi^{\alpha}_c\phi^{\alpha}_c}\xrightarrow{\gamma\to\infty}& 2\Delta^\text{Gauss}_{\phi^{\alpha}_c\phi^{\alpha}_c}-d/2 &\simeq 2-3\epsilon/2,\nonumber\\ \Delta^\text{WF}_{\phi^{\alpha}_c\phi^{\alpha}_c}\xrightarrow{\gamma\to\infty}& 2\Delta^\text{WF}_{\phi^{\alpha}_c\phi^{\alpha}_c}-d/2 &\simeq 2+\epsilon/2.
\end{eqnarray*}
As a result, at the IR nonunitary fixed point which can be reached from the Gaussian fixed point, one should expect to see a nontrivial Keldysh two-point function $\left\langle\mathcal{O}_{c}(x)\mathcal{O}_{c}(0)\right\rangle \sim x^{-\left(4-3\epsilon\right)}$;
at the UV nonunitary fixed point which is backshot from the WF fixed point, however, one should expect to see $\left\langle\mathcal{O}_{c}(x)\mathcal{O}_{c}(0)\right\rangle \sim x^{-\left(4+\epsilon\right)}$. Both two-point functions should be conformal invariant.

\section{Discussion and conclusions}
From a holographic perspective, two unitary IR and UV CFTs connected by a double-trace deformation $\mathbf{O}^\intercal\mathbf{O}$ of large $N$ have a natural explanation in the bulk AdS space: the only difference between them amounts to a change of the boundary conditions~\cite{deform-ads_kw99} to the bulk equations of motion (EOM) near $z\to0$ in the Poincar\'e patch,
\begin{eqnarray*}
	\label{eq_pp}
ds^2&=&\frac{1}{z^2}\left(dt^2-dz^2-\sum^{d-1}_{i=1}dx_i^2\right).
\end{eqnarray*}
In particular, the bulk fields $\mathbf{\Phi}=\left(\Phi_1,\Phi_2,\cdots\right)^\intercal$ dual to the boundary fields $\mathbf{O}=\left(\mathcal{O}_1,\mathcal{O}_2,\cdots\right)^\intercal$ of dimension $\Delta_1,\Delta_2,\cdots$ 
can be expanded near $z=\varepsilon\ll 1$ as $\Phi_i(z,\mathbf{x})\sim\alpha_i(\mathbf{x})\varepsilon^{d-\Delta_i}+\beta_i(\mathbf{x})\varepsilon^{\Delta_i}$~\cite{deform-ads-cft_w02}. 
Multipoint functions of $\mathcal{O}_i$ of the undeformed CFT can be recovered by taking $\alpha_i(\mathbf{x})\to0$. On the other hand, by taking $\beta_i(\mathbf{x})\to0$, one instead recovers multipoint functions for the deformed CFT located at the other end of the RG flow. The roles of $\alpha_i(\mathbf{x})$ and $\beta_i(\mathbf{x})$ are thus exchanged.

What is special for Lorentzian AdS space, 
however, is that the bulk EOM allows not one but two independent smooth solutions that vanish exponentially when $z\to\infty$~\cite{ads-cft-boundary_bklt99}. Therefore, unlike in Euclidean space where $\Phi_i(z,\mathbf{x})$ could be completely fixed by the EOM with a boundary condition $\Phi_i(\varepsilon,\mathbf{x})=\varepsilon^{d-\Delta_i}\Phi_i^{0}(\mathbf{x})$ near $\varepsilon=0$, 
here $\Phi_i(z,\mathbf{x})$ cannot be fixed. The ambiguity therein is nothing but exactly the time-ordering ambiguity of propagators in the corresponding Minkowski CFT~\cite{ads-cft-boundary_bklt99} and is usually fixed by hand to match known propagators in the Euclidean AdS space by Wick rotation~\cite{ads-cft-time-order_ss02}. However, there are no general rules (other than thermodynamic reasoning) forbidding us to fix the ambiguity in other ways. Thus, under the Keldysh formalism, $\mathcal{O}_{i,c}$ and $\mathcal{O}_{i,q}$ in the boundary theory
can still be coupled to the same boundary condition $\Phi_i^{0}(\mathbf{x})$ 
but mapped to two different solutions $\Phi_i(z,\mathbf{x})$ in the bulk theory.
Naturally, we want to know how we can reproduce our deformed Keldysh theory on the boundary by imposing some boundary condition to the bulk theory, a solution of which should require rewriting the bulk theory in the Keldysh formalism. We notice that similar interests have already appeared in recent ongoing studies~\cite{keldysh-ads-cft_gcl18,keldysh-ads-cft_jlr20}.

To conclude, we formalize a nonunitary Keldysh path integral that satisfies two ``\emph{CPT}'' constraints. 
We find that the introduced Lindblad term, which is responsible for driven-dissipative dynamics, has the natural form of a double-trace deformation $\mathcal{O}^2$. A calculation in the large $N$ limit shows the possible existence of a new, nonunitary, and nonthermal conformal fixed point induced by the double-trace deformation. 
This fixed point is IR when $\Delta<d/2$ but UV when $\Delta>d/2$, given $\Delta$ the scaling dimension of $\mathcal{O}$. Interestingly, such a fixed point remains physical in both IR and UV regimes in the large $N$ limit, suggesting that general driven-dissipative dynamics may unexpectedly differ from thermodynamics in its UV/IR relevance as well.  The retarded/advanced functions and the Keldysh function are also calculated at this nonunitary fixed point for general $\Delta$, as well as for two specific examples of a $\left(0+1\right)$-dimensional massive scalar field and a $\left(4-\epsilon\right)$-dimensional $O(N)$ massless scalar field. For the former example, the known feature of infinite heating in quantum collapse models is successfully recovered; for the latter, the RG flows between the two unitary Gaussian and WF fixed points and the two corresponding  nonunitary fixed points originated from them are studied.  We look forward to a thorough investigation of the possible AdS/CFT correspondence of our results in the bulk theory using the Keldysh formalism.

\acknowledgments
X.~M. thanks the support of Naoki Hanzawa and Houtarou Oreki. This work is supported by the NetSeed: Seedling Research Award of Northeastern University.

\appendix
\section{CONSTRUCT THE KELDYSH ACTION $S[\phi_{+},\phi_{-}]$ FROM THE TIME-EVOLUTION SUPEROPERATOR $\mathcal{V}(t^f,t^i)$}
\label{append_s-from-l}

Construction of $S[\phi_{+},\phi_{-}]$ from $\mathcal{V}(t^f,t^i)$ is similar to the traditional Feynman path integral approach. We stress that our approach of constructing $S[\phi_{+},\phi_{-}]$ is designed for \emph{relativistic} dynamics where the spacetime derivatives of $\phi_{\pm}$ are Lorentz covariant. This is different from the earlier coherent-state-based approach for nonrelativistic dynamics~\cite{keldysh_sbd16}. We start by rewriting the time-evolution superoperator $\mathcal{V}$ in a general exponential form, 
\begin{eqnarray}
\mathcal{V}(t^f,t^i)=\mathcal{T}e^{\int_{t^i}^{t^f}dt\mathcal{L}(t)},
\end{eqnarray}
where $\mathcal{T}$ is the anti-time-ordering superoperator. Dividing the time integral into $N$ equal intervals of length $\Delta t=(t^f-t^i)/N$ yields
\begin{eqnarray}
\mathcal{V}(t^f,t^i)\simeq e^{\Delta t \mathcal{L}(t_{N-1})}e^{\Delta t \mathcal{L}(t_{N-2})}\cdots e^{\Delta t \mathcal{L}(t_0)}
\end{eqnarray}
which is an exact equality when $N\to\infty$ by applying the Baker–Campbell–Hausdorff formula. Here the $n$th step is labeled by $t_n=t^i+n\Delta t$. Therefore, we can divide the Hilbert-space time evolution into infinitesimal segments,
\begin{eqnarray}
\label{eq_segments}
	&&\left\langle\phi_{+}^f(\mathbf{x})\right|\left[\mathcal{V}(t^f,t^i)(\left|\phi_{+}^i(\mathbf{x})\right\rangle\left\langle\phi_{-}^i(\mathbf{x})\right|)\right]\left|\phi_{-}^f(\mathbf{x})\right\rangle \nonumber\\
	&=&\int d\phi_{+,N-1}\int d\phi_{-,N-1}\cdots\int d\phi_{+,1}\int d\phi_{-,1}\nonumber\\
	&&\left\langle\phi_{+}^f\right|\left[e^{\Delta t \mathcal{L}(t_{N-1})}(\left|\phi_{+,N-1}\right\rangle\left\langle\phi_{-,N-1}\right|)\right]\left|\phi_{-}^f\right\rangle\nonumber\\
	&&\left\langle\phi_{+,N-1}\right|\left[e^{\Delta t \mathcal{L}(t_{N-2})}(\left|\phi_{+,N-2}\right\rangle\left\langle\phi_{-,N-2}\right|)\right]\left|\phi_{-,N-1}\right\rangle\nonumber\\
	&&\cdots\nonumber\\
	&&\left\langle\phi_{+,2}\right|\left[e^{\Delta t \mathcal{L}(t_{1})}(\left|\phi_{+,1}\right\rangle\left\langle\phi_{-,1}\right|)\right]\left|\phi_{-,2}\right\rangle\nonumber\\
	&&\left\langle\phi_{+,1}\right|\left[e^{\Delta t \mathcal{L}(t_{0})}(\left|\phi_{+}^i\right\rangle\left\langle\phi_{-}^i\right|)\right]\left|\phi_{-,1}\right\rangle,
\end{eqnarray}
which is valid provided that $\mathcal{L}(t_n)(\cdot)$ is a linear superoperator, i.e.,
\begin{eqnarray}
&&\int d\phi_{+,n}\int d\phi_{-,n} f(\phi_{+,n},\phi_{-,n})
\left[\mathcal{L}(t_{n})(\left|\phi_{+,n}\right\rangle\left\langle\phi_{-,n}\right|)\right]
\nonumber\\
&=&
\mathcal{L}(t_{n})(\int d\phi_{+,n}\int d\phi_{-,n} f(\phi_{+,n},\phi_{-,n})\left|\phi_{+,n}\right\rangle\left\langle\phi_{-,n}\right|)\nonumber
\end{eqnarray}
for arbitrary $f$. Here, each $\int d\phi_{\pm,n}$ is a shorthand of $\prod_{\mathbf{x}}\int d\phi_{\pm,n}(\mathbf{x})$, i.e., the product of integrals for every $\phi_{\pm,n}(\mathbf{x})$ living at $\mathbf{x}$ over the space (and same for $\int d\Pi_{\pm,n}$ below), and hence, $\int d\phi_{\pm,n} \left|\phi_{\pm,n}\right\rangle\left\langle\phi_{\pm,n}\right|
=1$. The space label $\mathbf{x}$ is also omitted. 

From Eq.~(\ref{eq_segments}), one can proceed to write down a path integral in terms of the bra and ket fields $\phi_{\pm}$ in the limit of $N\to \infty$. Note that complexification of $\phi_{\pm}$ can also be done by adding a new real scalar field $\varphi_{\pm}$, which together establishes an $SO(2)$ symmetry between $\phi_{\pm}$ and $\varphi_{\pm}$. In the following, we will look at both unitary and nonunitary dynamics and proceed to derive the corresponding Keldysh actions.

\subsection{Unitary dynamics}
For unitary QFTs which admit linear Hamiltonian dynamics, i.e., $\mathcal{V}(t^f,t^i)\rho(t^i)=e^{-i H (t^f-t^i)}\rho(t^i)e^{i H (t^f-t^i)}$, the time evolution reads
\begin{eqnarray}
\left\langle\phi_{+}^f(\mathbf{x})\right|\left[e^{-i H (t^f-t^i)}\left|\phi_{+}^i(\mathbf{x})\right\rangle\left\langle\phi_{-}^i(\mathbf{x})\right|e^{i H (t^f-t^i)}\right]\left|\phi_{-}^f(\mathbf{x})\right\rangle\nonumber
\end{eqnarray}
which is nothing but a twofold of the time evolution amplitude in terms of a single operator $\phi(\mathbf{x})$.
We further assume that the time-independent Hamiltonian $H$ can be written in terms of a local integral, $H=\int g(\mathbf{x}) \Pi^2(\mathbf{x})+v(\phi(\mathbf{x}))$ (for arbitrary $g$ and $v$), where $\Pi(\mathbf{x})$ as the conjugate momentum operator of $\phi(\mathbf{x})$ only appears in a quadratic form. Then, each infinitesimal segment in Eq.~(\ref{eq_segments}) can be rewritten as
\begin{eqnarray}
\label{eq_h-path-integral}
&&\left\langle\phi_{+,n+1}\right|\left[e^{-iH\Delta t }\left|\phi_{+,n}\right\rangle\left\langle\phi_{-,n}\right|e^{iH\Delta t}\right]\left|\phi_{-,n+1}\right\rangle\nonumber\\
&=&\int \frac{d\Pi_{+,n}}{2\pi}e^{-i\Pi_{+,n}\left(\phi_{+,n+1}-\phi_{+,n}\right)}e^{-iH_{+}\Delta t }\nonumber\\
&&\int \frac{d\Pi_{-,n}}{2\pi}e^{i\Pi_{-,n}\left(\phi_{-,n+1}-\phi_{-,n}\right)}e^{iH_{-}\Delta t },
\end{eqnarray}
where $H_{+}$ only depends on $\phi_{+}$ and $\Pi_{+}$ (similarly for $H_{-}$). In the second step of Eq.~(\ref{eq_h-path-integral}), we have used $\left\langle\Pi_{\pm}|\phi_{\pm}\right\rangle=\left(2\pi\right)^{1/2}e^{-i\Pi_{\pm}\phi_{\pm}}$.
Noticing that the integrand in Eq.~(\ref{eq_h-path-integral}) is an exponential of a quadratic term in $\Pi_{\pm,n}$, we can calculate the Gaussian integral exactly and find that Eq.~(\ref{eq_h-path-integral}) is proportional to 
\begin{eqnarray}
&&e^{i \left\{\left(g^{-1}/4\right) \left[\left({\phi_{+,n+1}-\phi_{+,n}}\right)/\Delta t\right]^2-v(\phi_{+,n})\right\}\Delta t }\nonumber\\
&&+e^{-i \left\{\left(g^{-1}/4\right) \left[\left({\phi_{-,n+1}-\phi_{-,n}}\right)/\Delta t\right]^2-v(\phi_{-,n})\right\}\Delta t }\nonumber\\
&=&e^{i \left(L(\phi_{+,n})-L(\phi_{-,n})\right)\Delta t }.
\end{eqnarray}
It is obvious that the Lagrangian $L(\phi_{\pm})$ and the Hamiltonian $H(\phi_{\pm})$ are connected by the Legendre transformation. From this Lagrangian expression, one can go back to Eq.~(\ref{eq_segments}), take $N\to \infty$, and construct the Keldysh action,
\begin{eqnarray}
\label{eq_keldysh_unitarity}
S[\phi_{+},\phi_{-}]=S[\phi_{+}]-S[\phi_{-}],
\end{eqnarray}
where $S[\phi_{\pm}]=\int_{t^i}^{t^f}dt\int d\mathbf{x}L(\phi_{\pm})$.

The unitary Keldysh action [Eq.~(\ref{eq_keldysh_unitarity})] is of special interest~\cite{keldysh-i_hlr17, keldysh-ii_hlr17}. We can see that it is just two copies of the ordinary unitary action. Since there is no crossing term between $\phi_{+}$ and $\phi_{-}$, it seems that $\phi_{+}$ and $\phi_{-}$ are completely independent. However, the correlations between them [as defined by Eq.~(\ref{eq_keldysh-z})] are not zero due to the upper-limit boundary condition imposed on Eq.~(\ref{eq_keldysh-z}). To see this, we look at two-point functions for the vacuum. The ground state $\left|0\right\rangle$ is picked up by taking $t^f-t^i\to (1-i\varepsilon)\infty$ with an infinitesimal Wick rotation applied. The initial condition can thus be safely taken as $\rho(t^i)=\left|0\rangle\langle 0\right|$, which no longer depends on $t^i$, thus recovering the translational symmetry in Eq.~(\ref{eq_keldysh-time-order}). Noticing the cyclic property of Eq.~(\ref{eq_keldysh-time-order}),  we have
\begin{eqnarray}
\label{eq_keldysh++--}
\left\langle\phi_{-}(x)\phi_{+}(0)\right\rangle
&=&\left\langle 0\right|\phi(x)\phi(0)\left|0\right\rangle,\nonumber\\ \left\langle\phi_{+}(x)\phi_{-}(0)\right\rangle
&=&\left\langle 0\right|\phi(0)\phi(x)\left|0\right\rangle,\nonumber\\
\left\langle\phi_{+}(x)\phi_{+}(0)\right\rangle
&=&\left\langle 0\right|\mathcal{T}\left\{\phi(x)\phi(0)\right\}\left|0\right\rangle,\nonumber\\
\left\langle\phi_{-}(x)\phi_{-}(0)\right\rangle
&=&\left\langle 0\right|\mathcal{\bar{T}}\left\{\phi(x)\phi(0)\right\}\left|0\right\rangle,
\end{eqnarray}
which cover all possible time orders made of two local operators. Perturbative diagrams thus can be drawn from these propagators.
Through a Fourier transform, we see that the correlation between $\phi_{+}$ and $\phi_{-}$ is not zero only when the exchange of particles $\phi$-$\phi$ is on shell~\cite{keldysh-renorm_bjlr17}. In addition, since there are no vertices mixing $\phi_{+}$ and $\phi_{-}$ in the unitary dynamics $S[\phi_{+}]-S[\phi_{-}]$, the first two in Eq.~(\ref{eq_keldysh++--}) are the only connection between $\phi_{+}$ and $\phi_{-}$ and are called \emph{cut propagators}~\cite{keldysh-renorm_bjlr17}---namely, any diagram under the Keldysh formalism will reduce to two independent Feynman diagrams for the bra and ket fields, respectively, after cutting these on-shell propagators.

By direct inspection, the sum of all correlation functions in Eq.~(\ref{eq_keldysh++--}) is always equal to zero, a result of the conservation of probability~\cite{keldysh_sbd16}.
This is why, after some rearrangement, in the Keldysh basis
$
\left\langle{\Phi}(x){\Phi}^\intercal(0)\right\rangle
=\begin{pmatrix}
\left\langle 0\right|\{\phi(x),\phi(0)\}\left|0\right\rangle & \theta(x)\left\langle 0\right|[\phi(x),\phi(0)]\left|0\right\rangle\\
\theta(-x)\left\langle 0\right|[\phi(0),\phi(x)]\left|0\right\rangle&0
\end{pmatrix}
$, 
the lower right corner is always equal to zero. Also, one sees
clearly that the upper right and the lower left corners are called the retarded and the advanced functions because of the existence of the Heaviside step function $\theta(x)$.

\subsection{Nonunitary dynamics}
We pick the simplest nonunitary and time-independent $\mathcal{L}$ that admits a Lindblad form [as in Eq.~(\ref{eq_master})] and satisfies the quantum master equation,
\begin{eqnarray}
\label{eq_simple-master}
&&\partial_t \rho=\mathcal{L}(\rho)=-i\left[H,\rho\right]-\left(\gamma/2\right)\left[\mathcal{O},\left[\mathcal{O},\rho\right]\right],
\end{eqnarray}
given an arbitrary real operator $\mathcal{O}$. The corresponding adjoint master equation~\cite{q-open-syst} is
\begin{eqnarray}
\label{eq_adjoint-master}
\partial_t A&=&\mathcal{L}^{\dagger}(A)=-i\left[A,H\right]-\left(\gamma/2\right)\left[\mathcal{O},\left[\mathcal{O},A\right]\right]
\end{eqnarray}
for an arbitrary operator $A$. Note that $\mathcal{L}$ is a linear superoperator, since
\begin{eqnarray}
\label{eq_lagrangian-expand}
&&\left\langle\phi_{+,n+1}\right|\left[\mathcal{L}(\left|\phi_{+,n}\right\rangle\left\langle\phi_{-,n}\right|)\right]\left|\phi_{-,n+1}\right\rangle\nonumber\\
&=&-i\left\langle\phi_{+,n+1}\right|H \left|\phi_{+,n}\right\rangle\left\langle\phi_{-,n}|\phi_{-,n+1}\right\rangle\nonumber\\
&&+i\left\langle\phi_{+,n+1}|\phi_{+,n}\right\rangle\left\langle\phi_{-,n}\right|H\left|\phi_{-,n+1}\right\rangle\nonumber\\
&&+\gamma\left\langle\phi_{+,n+1}\right|\mathcal{O} \left|\phi_{+,n}\right\rangle\left\langle\phi_{-,n}\right|\mathcal{O}\left|\phi_{-,n+1}\right\rangle\nonumber\\
&&-\left(\gamma/2\right)\left\langle\phi_{+,n+1}\right|\mathcal{O}^2 \left|\phi_{+,n}\right\rangle\left\langle\phi_{-,n}|\phi_{-,n+1}\right\rangle\nonumber\\
&&-\left(\gamma/2\right)\left\langle\phi_{+,n+1}|\phi_{+,n}\right\rangle\left\langle\phi_{-,n}\right|\mathcal{O}^2\left|\phi_{-,n+1}\right\rangle.\quad
\end{eqnarray}
If $\mathcal{O}$ does not contain spacetime derivatives of $\phi_{\pm}$, then for the last three terms on the right-hand side of Eq.~(\ref{eq_lagrangian-expand}), we can safely take
\begin{eqnarray}
\label{eq_overlap}
\left\langle\phi_{\pm,n+1}\right|\mathcal{O} \left|\phi_{\pm,n}\right\rangle\approx
\left\langle\phi_{\pm,n}\right|\mathcal{O} \left|\phi_{\pm,n}\right\rangle=\mathcal{O}_{\pm}(\phi_{\pm}),\qquad
\end{eqnarray}
where $\mathcal{O}_{\pm}$ are the bra and ket fields corresponding to $\mathcal{O}$. They appear in the following term:
\begin{eqnarray}
 -i\int dx \left(\mathcal{O}_{+}(\phi_{+})\mathcal{O}_{-}(\phi_{-})-\frac{1}{2}\mathcal{O}_{+}^2(\phi_{+})-\frac{1}{2}\mathcal{O}_{-}^2(\phi_{-})\right),\nonumber
\end{eqnarray}
which should be added to Eq.~(\ref{eq_keldysh_unitarity}) to yield the final nonunitary $S[\phi_{+},\phi_{-}]$ that matches Eq.~(\ref{eq_simple-master}). 

If $\mathcal{O}$ contains spacetime derivatives of $\phi_{\pm}$, then we have to take into account the nontrivial overlap between $\left|\phi_{\pm,n+1}\right\rangle$ and $\left|\phi_{\pm,n}\right\rangle$ and thus cannot invoke Eq.~(\ref{eq_overlap}). However, Matthews's theorem~\cite{matthews-theor_m49,matthews-theor-first-order_bd75,matthews-theor-high-order_k94} implies that we can safely neglect this complication and simply assume Eq.~(\ref{eq_overlap}) is valid for perturbatively calculating the multipoint functions.
To be more specific, Matthews's theorem has established a one-to-one correspondence between the Feynman path-integral formalism and the Hamiltonian dynamics in the perturbative regime (even when the
momentum dependence of the Hamiltonian is higher than second order~\cite{matthews-theor-high-order_k94}). Since the Keldysh formalism is nothing but just a double copy of the usual path-integral formalism, we expect that this correspondence holds here too. 

\section{DOUBLE-TRACE DEFORMATION}
\label{append_deform}
The usual approach of studying double-trace deformation~\cite{deform-irrelevant_py16, deform_gk03} of a Minkowski QFT is to deform a Euclidean QFT first and then apply the Wick rotation to clear up the time-ordering ambiguity. In the Keldysh formalism, however, the deformation coupling strength parameter is already complex, which encodes the time-ordering information. Therefore, one must work in the Minkowski space directly.

\subsection{Singlet}
We start with double-trace deformation of a singlet that contains only one double-trace scalar term $f \mathcal{O}^2$. In general, the partition function $Z_f[J]$ with source $J$ that describes the deformed QFT (which may not be unitary) is given by
\begin{eqnarray}
\label{eq_zf}
Z_f[J]&\equiv&e^{W_f[J]}=\left\langle e^{i\int d^d x\left(-\frac{f}{2}\mathcal{O}^2(x)+J(x)\mathcal{O}(x)\right)} \right\rangle_0,\quad
\end{eqnarray}
where $W_f[J]$ is the free energy, and $\left\langle\cdots \right\rangle_0\equiv\int D\phi \left(\cdots\right) \exp iS_\text{0}[\phi]$ is the path-integral measure for the undeformed QFT~\cite{deform_gk03}. By applying the Hubbard-Stratonovich transformation, one introduces an auxiliary field $\sigma$ to Eq.~(\ref{eq_zf}) and gets
\begin{eqnarray}
\label{eq_zf_hs}
Z_f[J]&=&\int D\sigma \left\langle e^{i\int d^d x\left[\frac{1}{2f}\sigma^2(x)+\left(\sigma+J\right)(x)\mathcal{O}(x)\right]} \right\rangle_0\nonumber\\
&=&\int D\sigma  e^{i\int d^d x \frac{1}{2f}\sigma^2(x)} e^{W_0[\sigma+J]},
\end{eqnarray}
where $W_0\equiv W_{f=0}$ is the undeformed free energy. Such a transformation is possible in the Minkowski space if and only if $\Im\{f\}<0$.
Note that under a change of variables $\tilde{\sigma}=\sigma+J$ and $\tilde{J}=J/f$, one has
\begin{eqnarray}
\label{eq_zf_legendre-trans}
e^{W_{\infty}[\tilde{J}]}&=&\lim\limits_{f\to\infty}\int D\tilde{\sigma}  e^{i\int d^d x\frac{1}{2f}\left(\tilde{\sigma}-f \tilde{J}\right)^2(x)} e^{W_0[\tilde{\sigma}]}\nonumber\\
&\propto&\int D\tilde{\sigma} e^{-i\int d^dx \tilde{\sigma}(x)\tilde{J}(x)+W_0[\tilde{\sigma}]},
\end{eqnarray}
up to a divergent multiplicative coefficient (contact term) $\exp (i f \int d^dx  \tilde{J}^2(x)/2)$. From Eq.~(\ref{eq_zf_legendre-trans}), one immediately sees that
the two free energies at the UV and IR fixed points, $W_{\infty}$ and $W_{0}$, are connected by the Legendre transformation~\cite{deform_gk03}.

If we assume that $\mathcal{O}$ after normalization is proportional to a hidden factor of $N^{-1}$ [e.g., $\mathcal{O}=\phi^{\alpha}\phi^{\alpha}$ for $O(N)$ model which after normalization will be divided by $N$], then $W_0$ admits a cumulant expansion,
\begin{eqnarray}
\label{eq_zf_large-n}
{W_{0}[\tilde{\sigma}]}&\simeq&-\frac{1}{2}\left\langle \left(\int d^d x  \tilde{\sigma}(x)\mathcal{O}(x)\right)^2\right\rangle_0+\cdots\nonumber\\
&=&-\frac{1}{2}\int d^d x \int d^d y \tilde{\sigma}(x)\tilde{\sigma}(y)\left\langle\mathcal{O}(x)\mathcal{O}(y)\right\rangle_0+\cdots,\nonumber\\
\end{eqnarray}
where higher terms are suppressed by factors of $N$. 

At the $O(N^0)$ level, only the first term which is quadratic in $\sigma$ in Eq.~(\ref{eq_zf_large-n}) is kept, where the undeformed two-point function is given by
$\left\langle\mathcal{O}(x)\mathcal{O}(y)\right\rangle_0=\int \frac{d^d k}{\left(2\pi\right)^d}e^{-ik\left(x-y\right)}G_0(k^2)$, with $G_0(k^2)$ the free propagator in the momentum space. Putting Eq.~(\ref{eq_zf_large-n}) into Eq.~(\ref{eq_zf_hs}), applying the Fourier transform, and then integrating out $\sigma(k)$ yield
\begin{eqnarray}
\label{eq_zf_final}
{W_{f}[J]}&=&\frac{i}{2}\int \frac{d^d k}{\left(2\pi\right)^d}J(k)\frac{iG_0(k^2)}{1+if G_0(k^2)}J^*(k),
\end{eqnarray}
plus some contact terms. Therefore, the full propagator is given by 
\[G(k^2)=-2 \delta^2 W_{f}[J]/\left(\delta J(k) \delta J^*(k)\right)=\left(G_0^{-1}(k^2)+if\right)^{-1}.\]
We see that $if$ is directly added to the two-point vertex function $G_0^{-1}(k^2)$ of the QFT. This is because in the large $N$ limit, a double-trace deformation is plainly additive to the effective action~\cite{deform-irrelevant_py16}.

\subsection{Multiplet}
Following similar procedures, Eq.~(\ref{eq_zf_final}) can be generalized to a multiplet where the deformation coupling strength is given by a matrix $\mathbf{f}$~\cite{deform-multiplet_bbdpr16}. The result is
\begin{eqnarray}
\mathbf{G}\simeq\left(\mathbf{G}_0^{-1}+i\mathbf{f}\right)^{-1}+O(N^1),
\end{eqnarray}
which leads to Eq.~(\ref{eq_g22}).

\section{HAMILTONIAN DYNAMICS OF ONE-DIMENSIONAL QUANTUM HARMONIC OSCILLATOR UNDER $q$ MEASUREMENT}
\label{append_qho}
Solving the adjoint master equation Eq.~(\ref{eq_adjoint-master}) with $H=\left(p^2+m^2q^2\right)/2$, $\mathcal{O}=q$, and $A=q,p,q^2,p^2,\cdots$, under the Gaussian approximation~\cite{contin-meas_js06} we get (given $[q,p]=i$)
\begin{eqnarray*}
	\partial_t \left\langle q(t) \right\rangle&=& \left\langle p(t) \right\rangle,\nonumber\\
	\partial_t \left\langle p(t) \right\rangle&=& -m^2 \left\langle q(t) \right\rangle,\nonumber\\
	\partial_t \left\langle q^2(t)\right\rangle_c&=& \left\langle q(t)p(t)+p(t)q(t) \right\rangle_c,\nonumber\\
	\partial_t \left\langle p^2(t)\right\rangle_c&=& \gamma-m^2\left\langle q(t)p(t)+p(t)q(t) \right\rangle_c,\nonumber\\
	\partial_t \left\langle q(t)p(t)+p(t)q(t) \right\rangle_c&=&2\left\langle p^2(t)\right\rangle_c-2 m^2 \left\langle q^2(t)\right\rangle_c,\nonumber
\end{eqnarray*}
where $\left\langle A B\right\rangle_c= \left\langle A B\right\rangle-\left\langle A\right\rangle\left\langle B\right\rangle$. We see that the only direct contribution of $\gamma$ is to $\left\langle p^2(t)\right\rangle_c$~\cite{contin-meas_js06}. This contribution generates momentum $p$ diffusion and heats the system at rate $\gamma$, leading to the long-time limit $\left\langle p^2(t)\right\rangle_c\sim \left\langle q^2(t)\right\rangle_c \sim \gamma t/2$ and leaving us an infinitely heated ``stationary'' state.

We can also calculate the correlations for $q$,
\begin{eqnarray*}
\label{eq_1dho++--}
\left\langle q_{+}(t) q_{+}(0)\right\rangle&=&-\left\langle q_{-}(t) q_{-}(0)\right\rangle\nonumber\\
&=&i\int \frac{d \omega}{2\pi}\frac{\omega^2-m^2-i\gamma}{\left(\omega^2-m^2\right)^2}e^{-i\omega t},\nonumber\\
\left\langle q_{+}(t) q_{-}(0)\right\rangle&=&\left\langle q_{-}(t) q_{+}(0)\right\rangle\nonumber\\
&=&i\int \frac{d \omega}{2\pi}\frac{-i\gamma}{\left(\omega^2-m^2\right)^2}e^{-i\omega t},
\end{eqnarray*}
from which we can immediately derive the linear response functions and see that they are independent of $\gamma$.

As for the Keldysh function $G^{K}$,
we may simply
write $G^{K}_{\mathcal{L}}=-2\gamma G^{R}G^{A}= 2\gamma/\left(\omega^2-m^2\right)^2$. But its Fourier transform diverges when $t\to \pm \infty$. Instead, noticing that 
\begin{eqnarray*}
&& \int \frac{d \omega}{2\pi} \frac{2\gamma}{\left[\left(\omega+i \varepsilon\right)^2-m^2\right]\left[\left(\omega-i \varepsilon\right)^2-m^2\right]} e^{-i\omega t}\nonumber\\
&=& \frac{\gamma}{4 m \varepsilon }\left\{\theta (t) \left[\frac{e^{-t \left(\varepsilon +i m\right)}}{m-i \varepsilon }+\frac{e^{-t \left(\varepsilon -i m\right)}}{m+i \varepsilon }\right] + \left(t\leftrightarrow -t\right)\right\}\nonumber\\
&\simeq& \frac{\gamma \cos (mt)}{2 m^2 \varepsilon }= \int \frac{d \omega}{2\pi}  \frac{\gamma}{m\varepsilon}\pi \delta\left(\omega^2-m^2\right) e^{-i\omega t},
\end{eqnarray*}
the true $G^{K}_{\mathcal{L}}$ should be given by $\left({m\varepsilon}\right)^{-1}{\gamma}\pi \delta\left(\omega^2-m^2\right)$, the Fourier transform of which does not diverge when $t\to \pm \infty$.

\section{DIAGRAMMATIC PERTURBATION FOR THE KELDYSH FORMALISM}
\label{append_diagram}
\begin{figure}[t!]
		\centering	
		{\includegraphics[width=35mm]{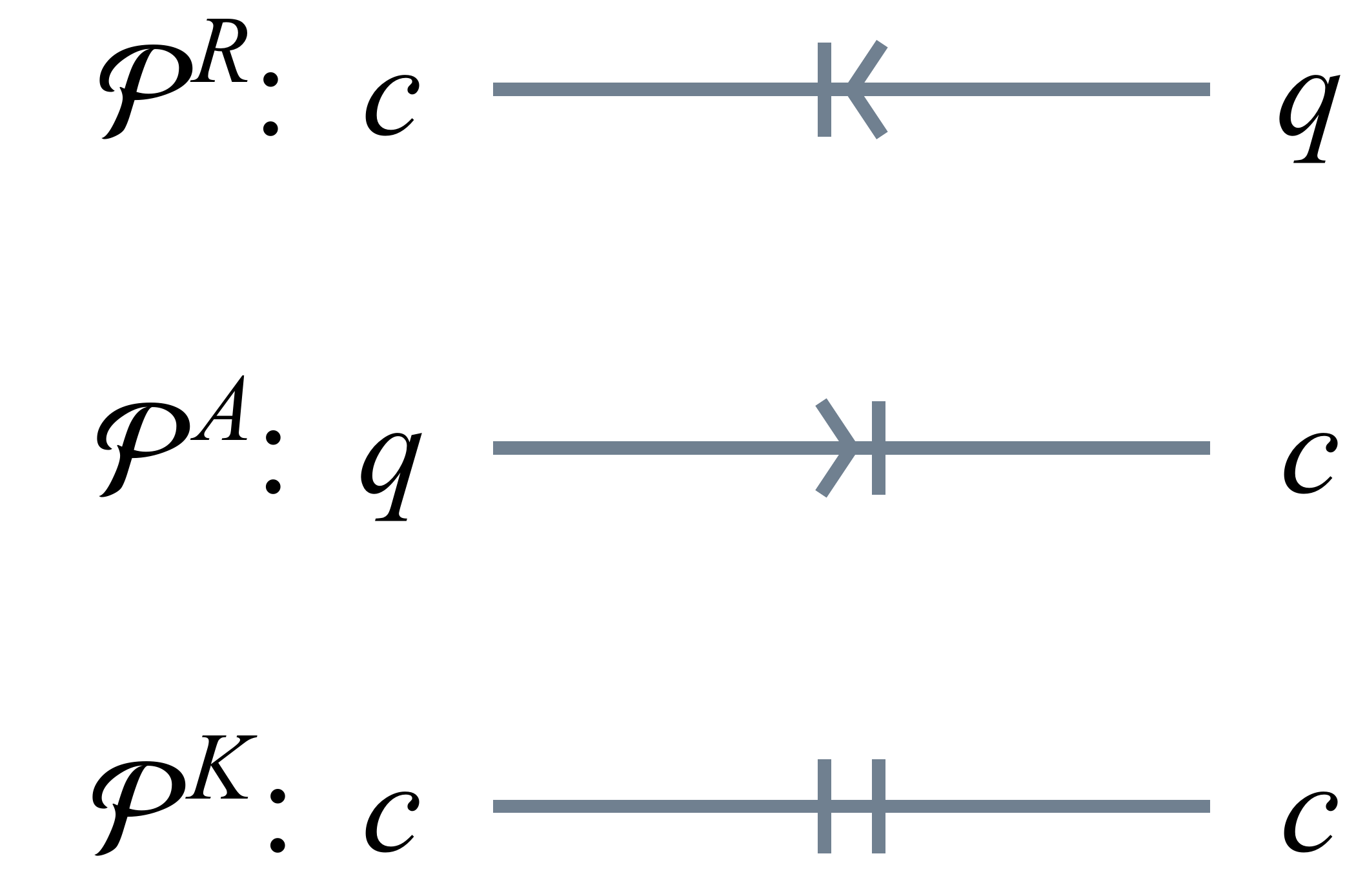}
			\vspace{0mm}
		}
	\caption{\label{fig_propagator}Free propagators in the Keldysh basis.\hfill\hfill}
\end{figure}

As a path-integral formalism, the Keldysh formalism naturally admits diagrammatic perturbative calculations. Here, we explain how to do the calculations for our second example, i.e., a $\left(4-\epsilon\right)$-dimensional $O(N)$ massless scalar field. The Lagrangian Eq.~(\ref{eq_nscalar-lagrangian}), after rewritten in the Keldysh basis, reads
\begin{eqnarray}
\label{eq_nscalar-lagrangian-cq}
&& L(\phi^{\alpha}_{c},\phi^{\alpha}_{q})=\partial_{\mu} \phi^{\alpha}_c\partial^{\mu} \phi^{\alpha}_q-m^2 \phi^{\alpha}_c \phi^{\alpha}_q+\frac{ic}{2}\phi^{\alpha}_c\phi^{\alpha}_c
\nonumber\\
&-&\frac{\lambda}{2\cdot 3!}\left({\phi^{\alpha}_c\phi^{\alpha}_c\phi^{\beta}_c\phi^{\beta}_q}+{\phi^{\alpha}_c\phi^{\alpha}_q\phi^{\beta}_q\phi^{\beta}_q}\right)
+2 i \gamma \left(\phi^{\alpha}_c \phi^{\alpha}_q\right)^2,\quad
\end{eqnarray}
where we have already included the quadratic terms $\phi^{\alpha}_c \phi^{\alpha}_q$ and $\phi^{\alpha}_c \phi^{\alpha}_c$. Note that $c$ should always be zero if we want the Keldysh formalism to be physical. At $c=0$, we have three propagators in total, 
\begin{eqnarray*}
\boldsymbol{\mathcal{P}}(k)=\begin{pmatrix}
	\mathcal{P}^{K}(k) & \mathcal{P}^{R}(k)\\
	\mathcal{P}^{A}(k)&0
\end{pmatrix},
\end{eqnarray*}
given by
\begin{eqnarray}
\label{eq_nscalar-propagator}
\mathcal{P}^{R/A}(\omega,\mathbf{k})&=&\frac{i}{\left(\omega\pm i\varepsilon\right)^2-\mathbf{k}^2-m^2},\nonumber\\
\mathcal{P}^{K}(\omega,\mathbf{k})&=&2\pi \delta (\omega^2-\mathbf{k}^2-m^2)
\end{eqnarray}
which can be drawn as shown in Fig.~\ref{fig_propagator}, following the same diagrammatic convention as in Ref.~\cite{keldysh-renorm_bjlr17}. Also, we see that the FDR $\mathcal{P}^{K}=\text{sign}(\omega)\left(\mathcal{P}^{R}-\mathcal{P}^{A}\right)$ is satisfied.

\begin{figure}[t!]
	\centering	
	\begin{minipage}[b]{40mm}
		{\includegraphics[width=40mm]{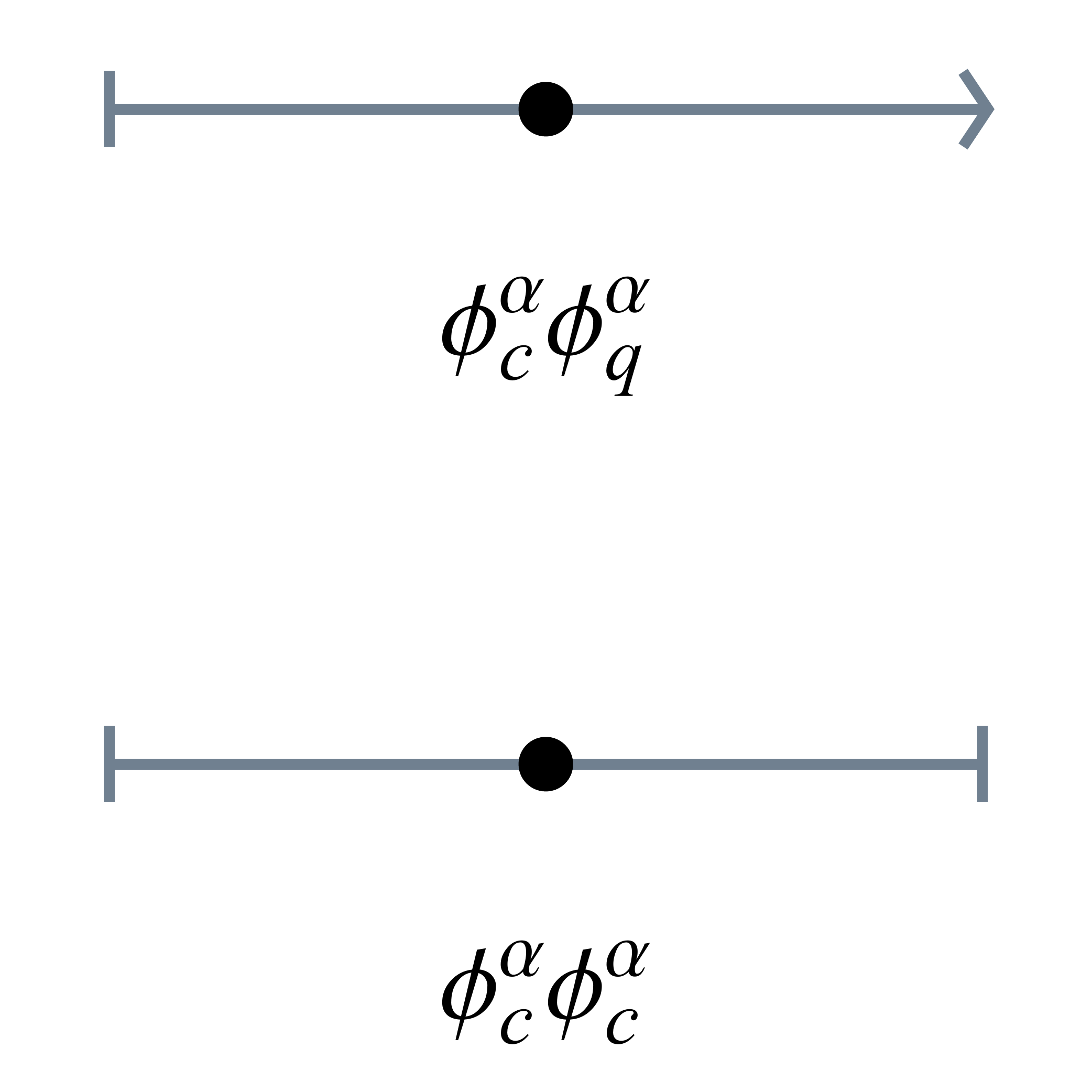}
			\vspace{6mm}
			\subcaption{\label{fig_vertex1}}}
	\end{minipage}
	\hspace{2mm}
	\begin{minipage}[b]{40mm}
		{\includegraphics[width=40mm]{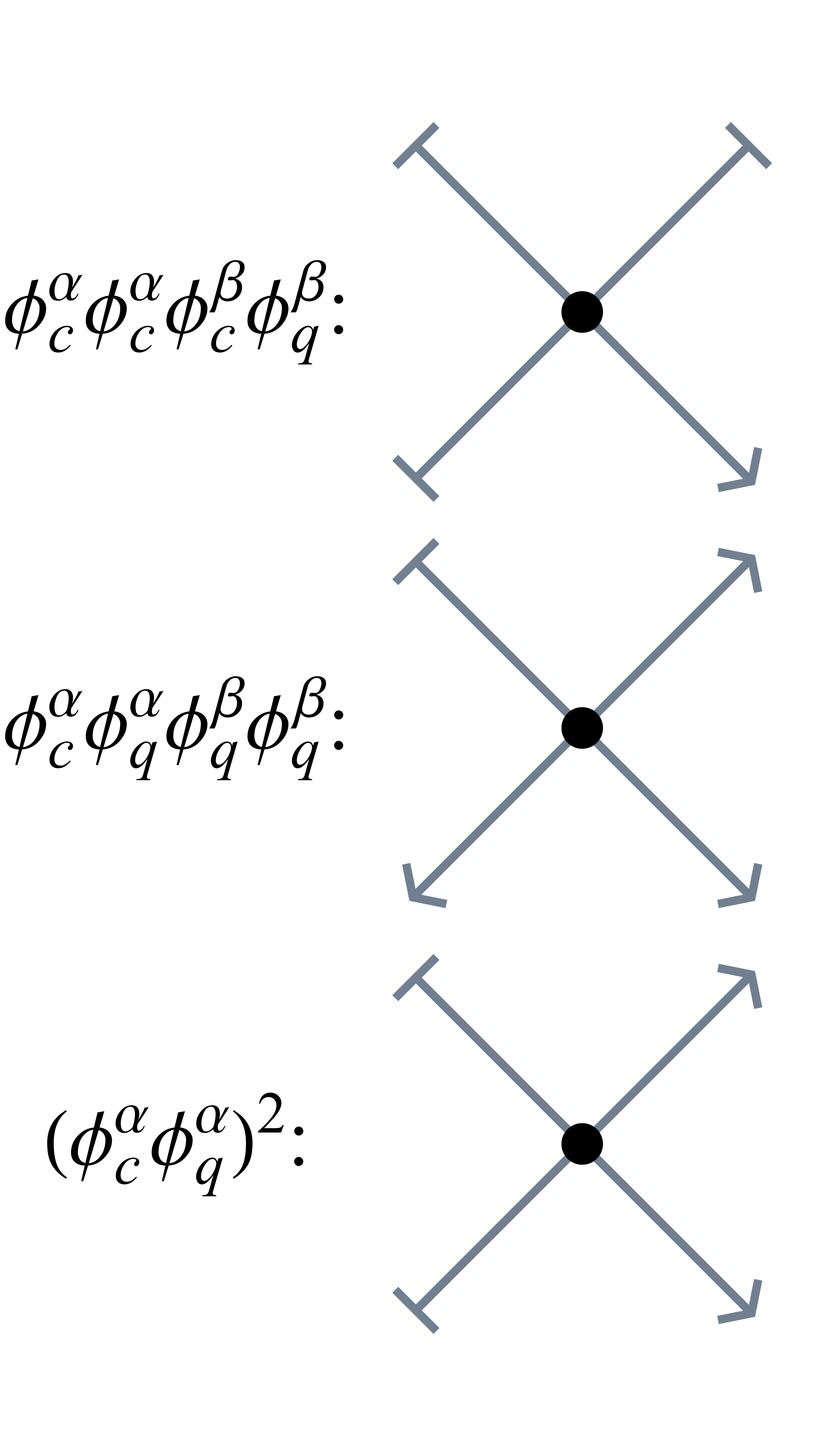}
			\subcaption{\label{fig_vertex2}}}
	\end{minipage}
	
	\caption{\label{fig_vertex}Vertices from Eq.~(\ref{eq_nscalar-lagrangian-cq}) in the Keldysh basis.\hfill\hfill}
\end{figure}

The vertices of which the beta functions we are interested in are shown in Fig.~\ref{fig_vertex}. Near the unitary WF fixed point, the beta functions for ${\phi^{\alpha}_c\phi^{\alpha}_c\phi^{\beta}_c\phi^{\beta}_q}$ and ${\phi^{\alpha}_c\phi^{\alpha}_q\phi^{\beta}_q\phi^{\beta}_q}$ should behave similarly, so we will only look at the renormalization of
one of the two vertices (${\phi^{\alpha}_c\phi^{\alpha}_c\phi^{\beta}_c\phi^{\beta}_q}$ only).

\begin{figure}[t!]
	\centering	
	\begin{minipage}[b]{40mm}
		{\includegraphics[width=40mm]{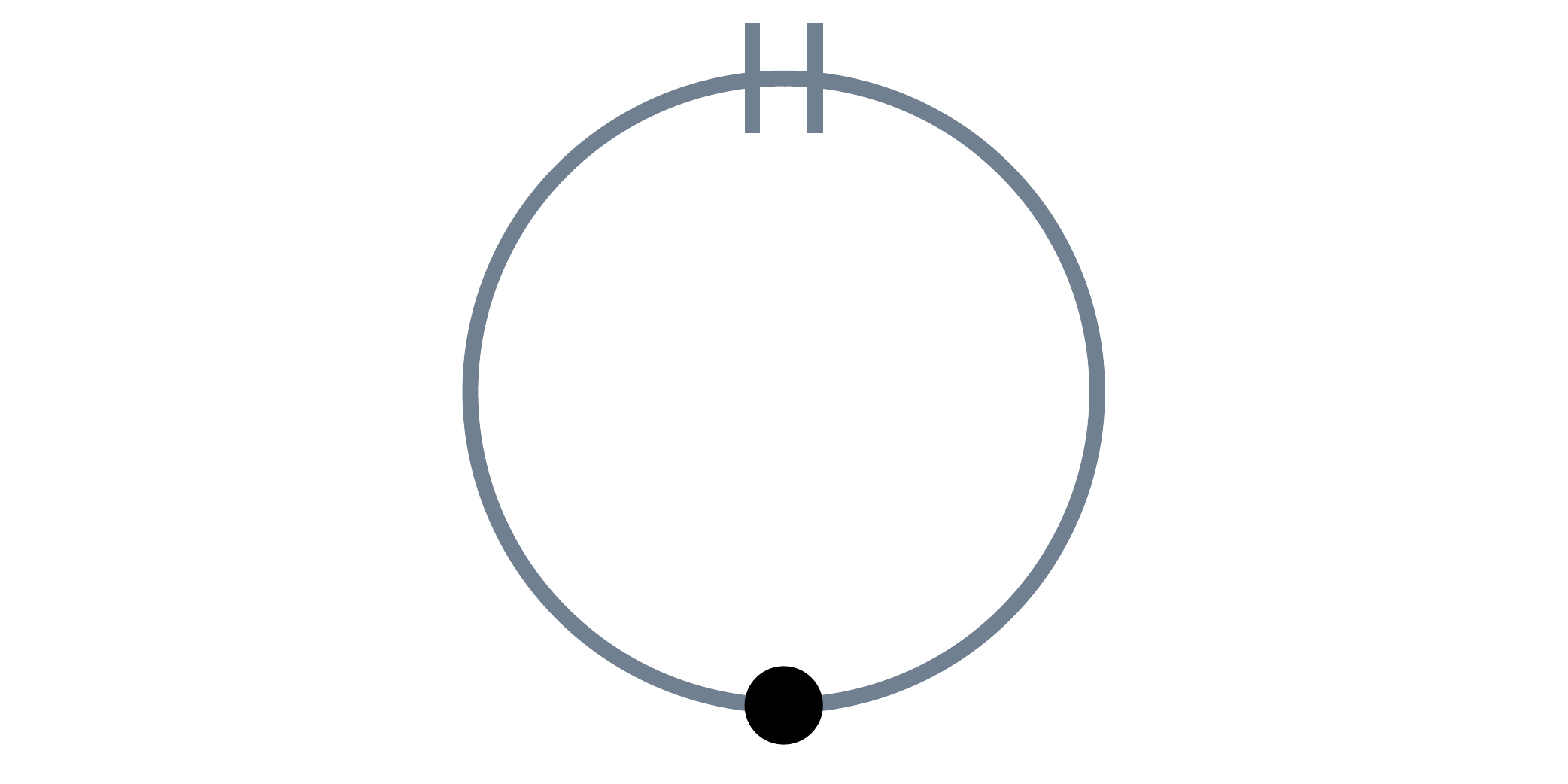}
			\subcaption{\label{fig_loop1}}}
	\end{minipage}
	\hspace{4mm}
	\begin{minipage}[b]{40mm}
		{\includegraphics[width=40mm]{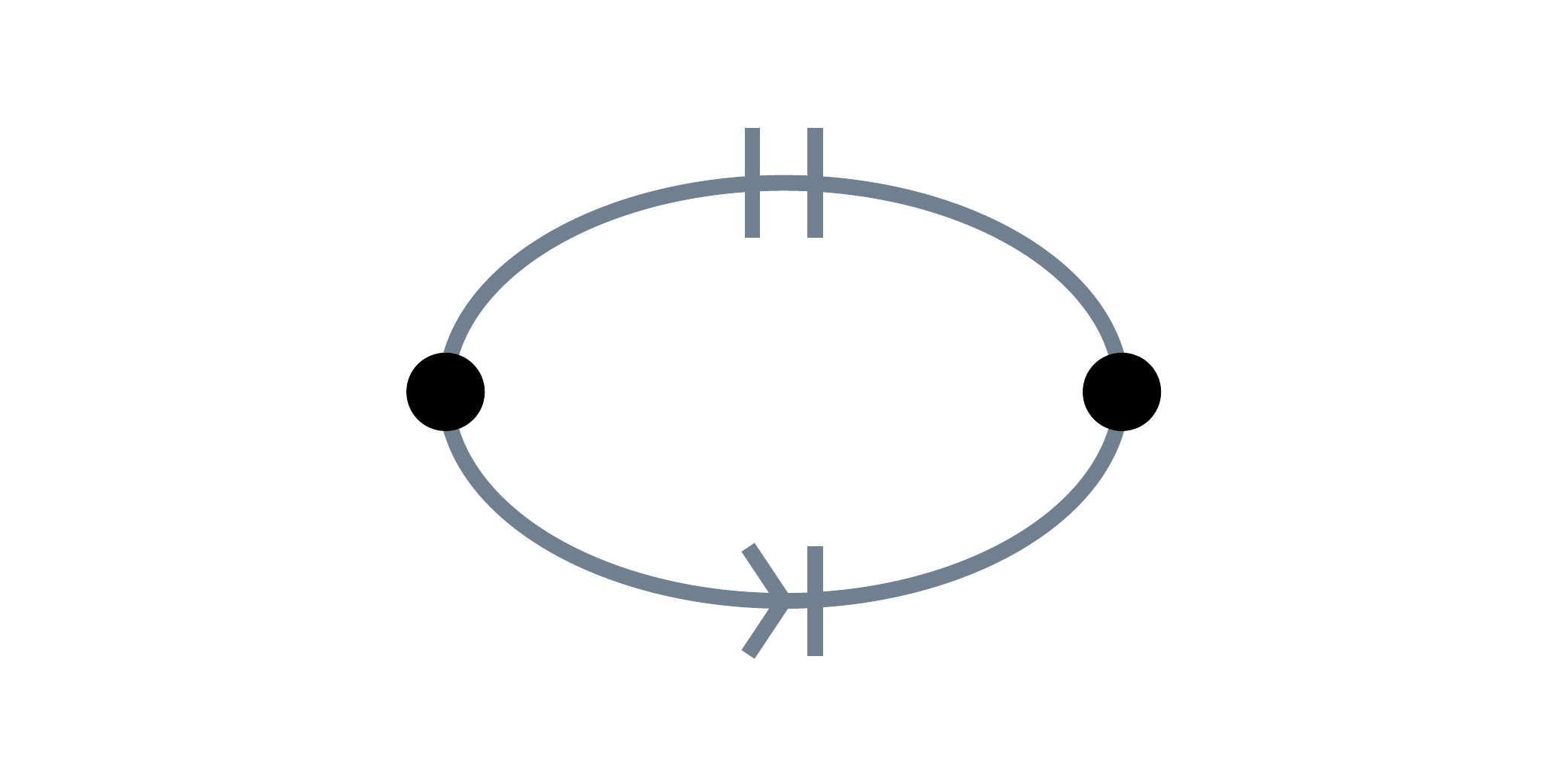}
			\subcaption{\label{fig_loop2}}}
	\end{minipage}
	\caption{\label{fig_loop}Divergent loops in the Keldysh basis at $O(c^0)$.\hfill\hfill}
\end{figure}
\begin{figure}[t!]
	\hspace{-12mm}
	\begin{minipage}[b]{35mm}
		{\includegraphics[width=30mm]{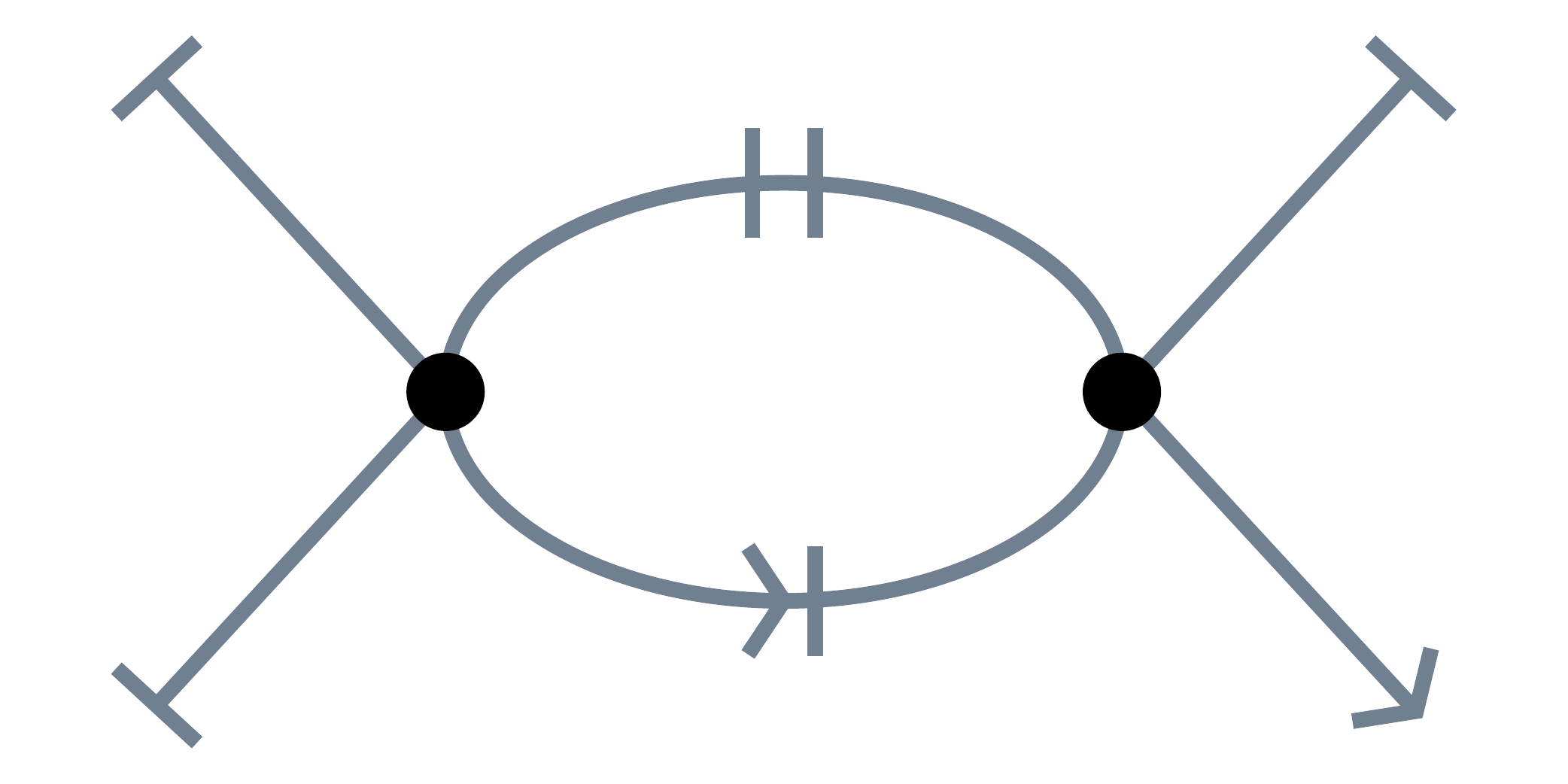}
			\vspace{-2mm}
			\subcaption{\label{fig_cccq-s}}}
	\end{minipage}
	\hspace{1mm}
	\begin{minipage}[b]{20mm}
		{\includegraphics[height=30mm]{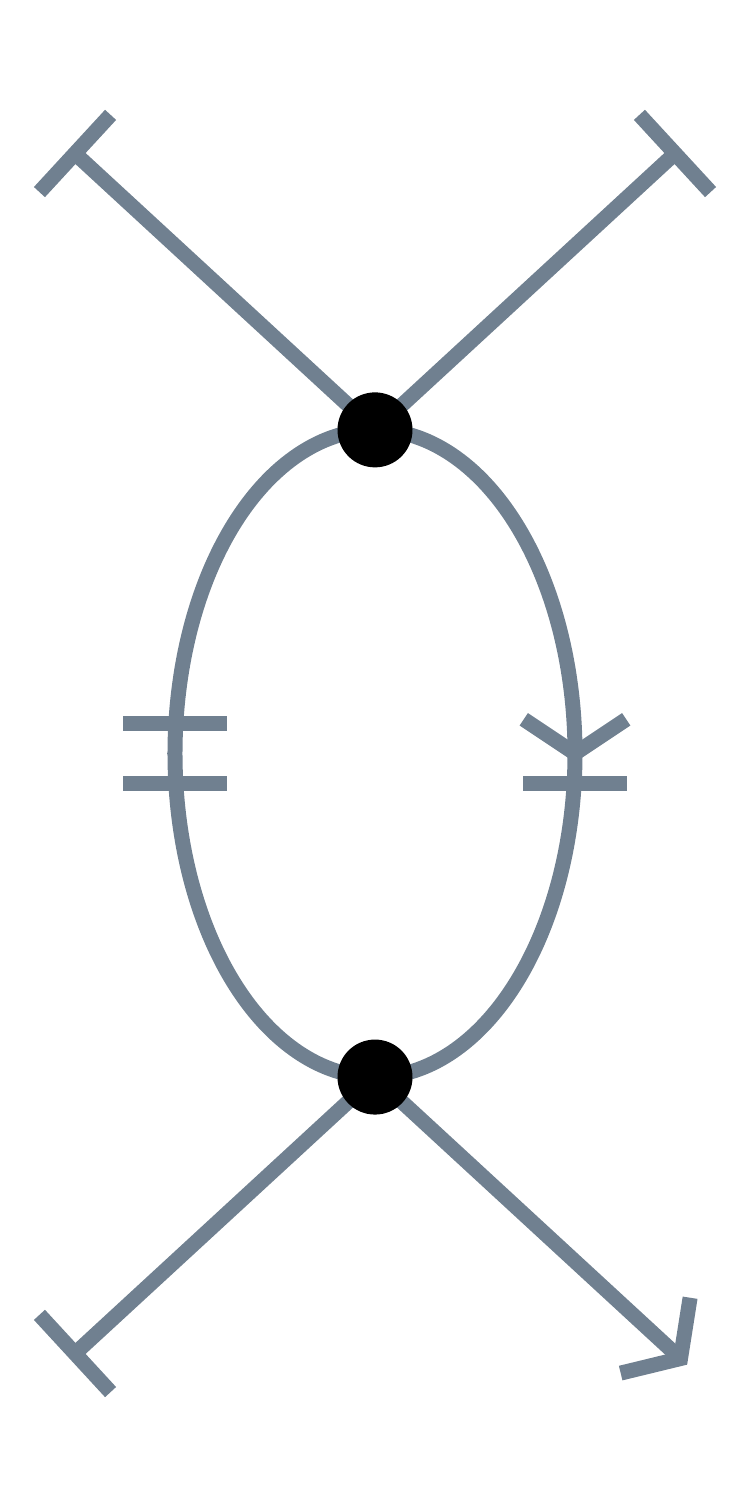}
			\subcaption{\label{fig_cccq-t}}}
	\end{minipage}
	\begin{minipage}[b]{20mm}
		{\includegraphics[height=30mm]{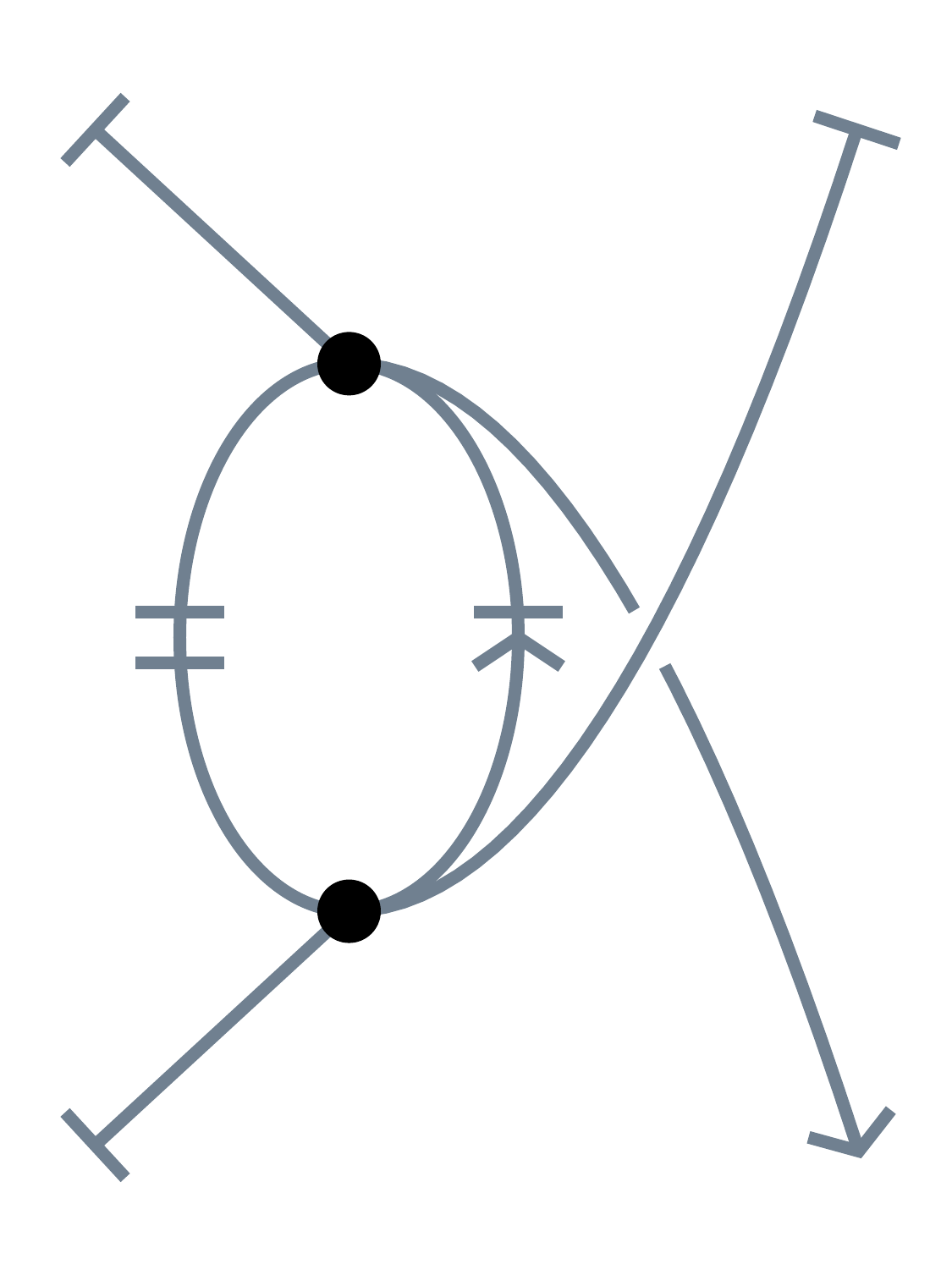}
			\subcaption{\label{fig_cccq-u}}}
	\end{minipage}
	\caption{\label{fig_cccq}One-loop corrections for the vertex ${\phi^{\alpha}_c\phi^{\alpha}_c\phi^{\beta}_c\phi^{\beta}_q}$.\hfill\hfill}
\end{figure}

\begin{figure}[t!]
	\hspace{-12mm}
	\begin{minipage}[b]{35mm}
		{\includegraphics[width=30mm]{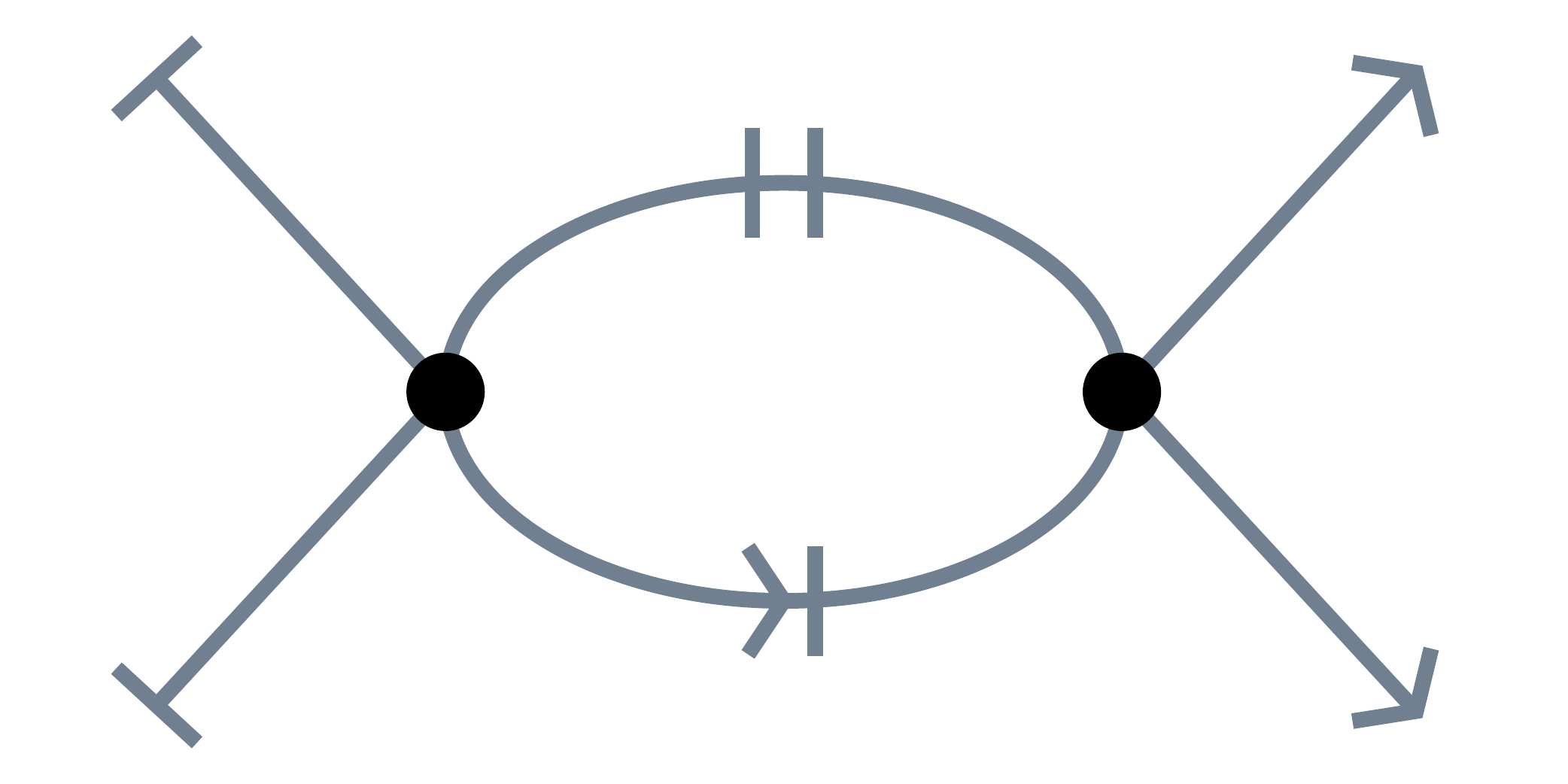}
			\vspace{-2mm}
			\subcaption{\label{fig_ccqq-s}}}
	\end{minipage}
	\begin{minipage}[b]{20mm}
		{\includegraphics[height=30mm]{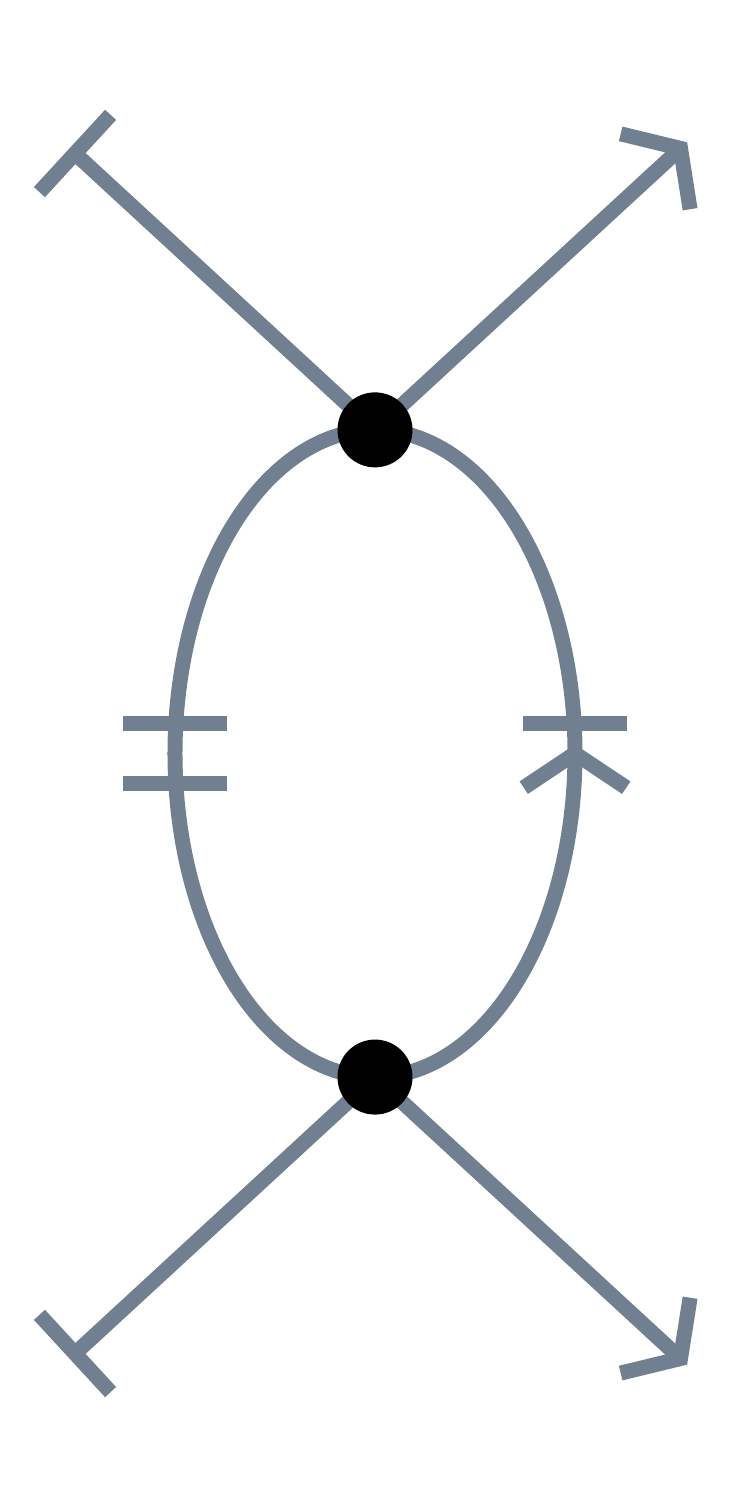}
			\subcaption{\label{fig_ccqq-t1}}}
	\end{minipage}
	\begin{minipage}[b]{20mm}
		{\includegraphics[height=30mm]{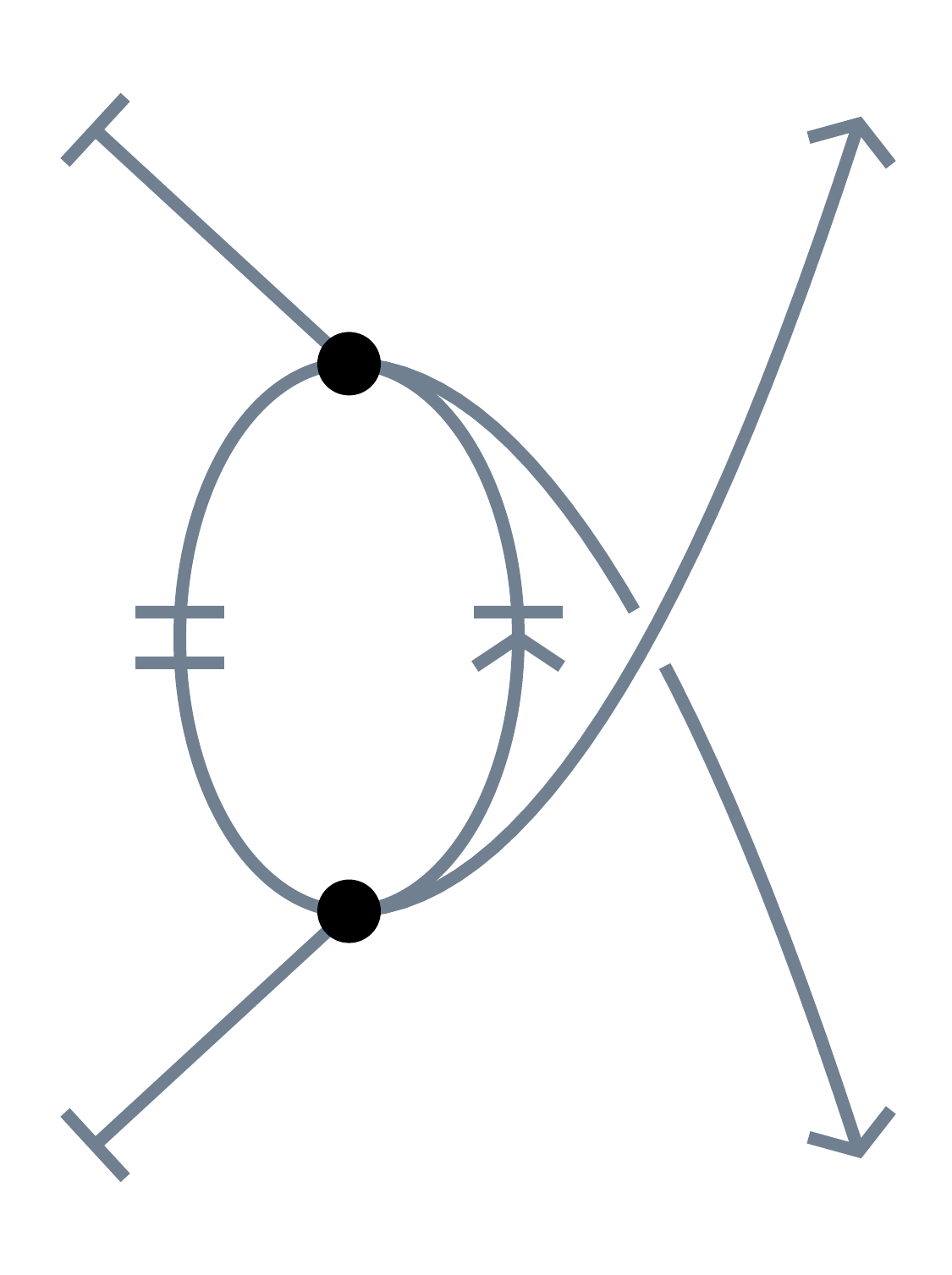}
			\subcaption{\label{fig_ccqq-u1}}}
	\end{minipage}
	\begin{minipage}[b]{20mm}
		{\includegraphics[height=30mm]{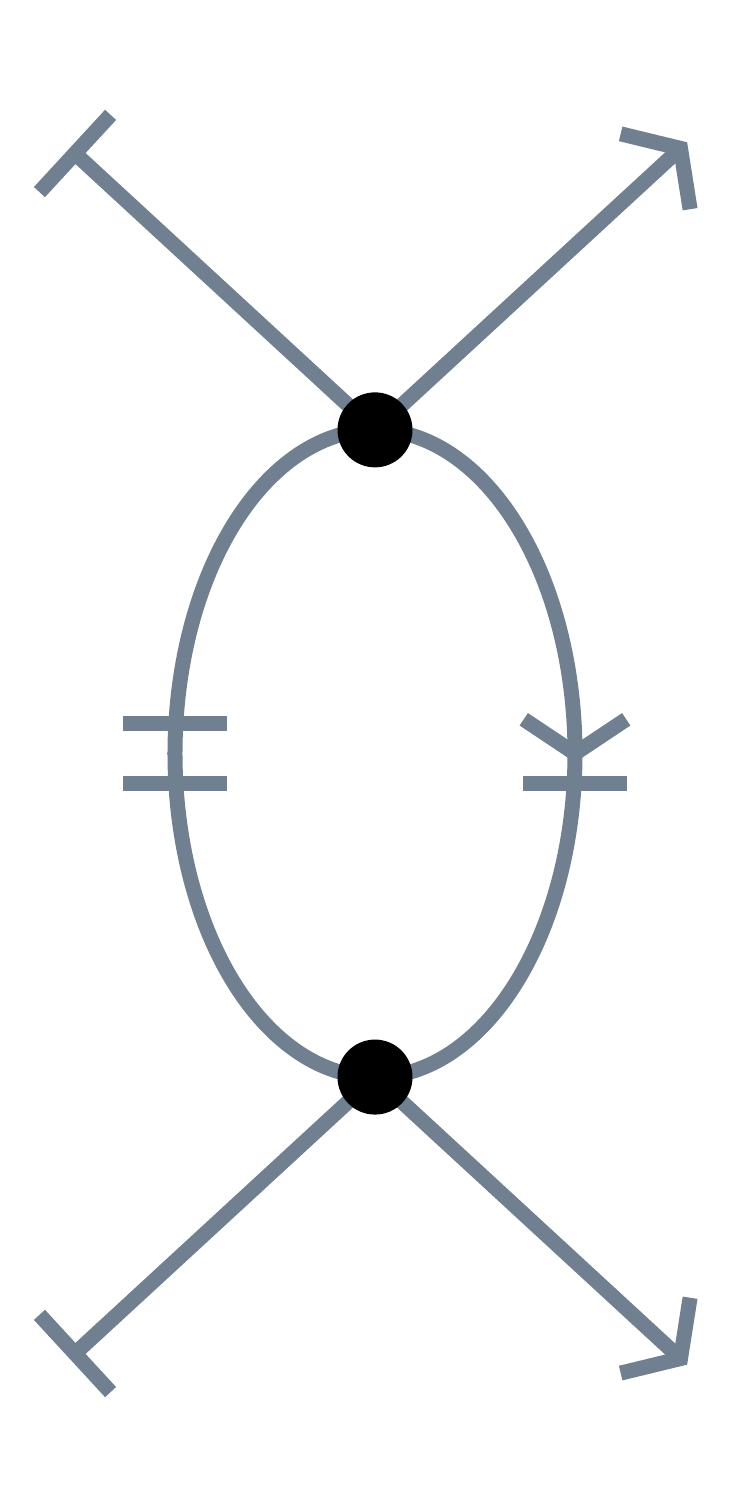}
			\subcaption{\label{fig_ccqq-t2}}}
	\end{minipage}
	\begin{minipage}[b]{20mm}
		{\includegraphics[height=30mm]{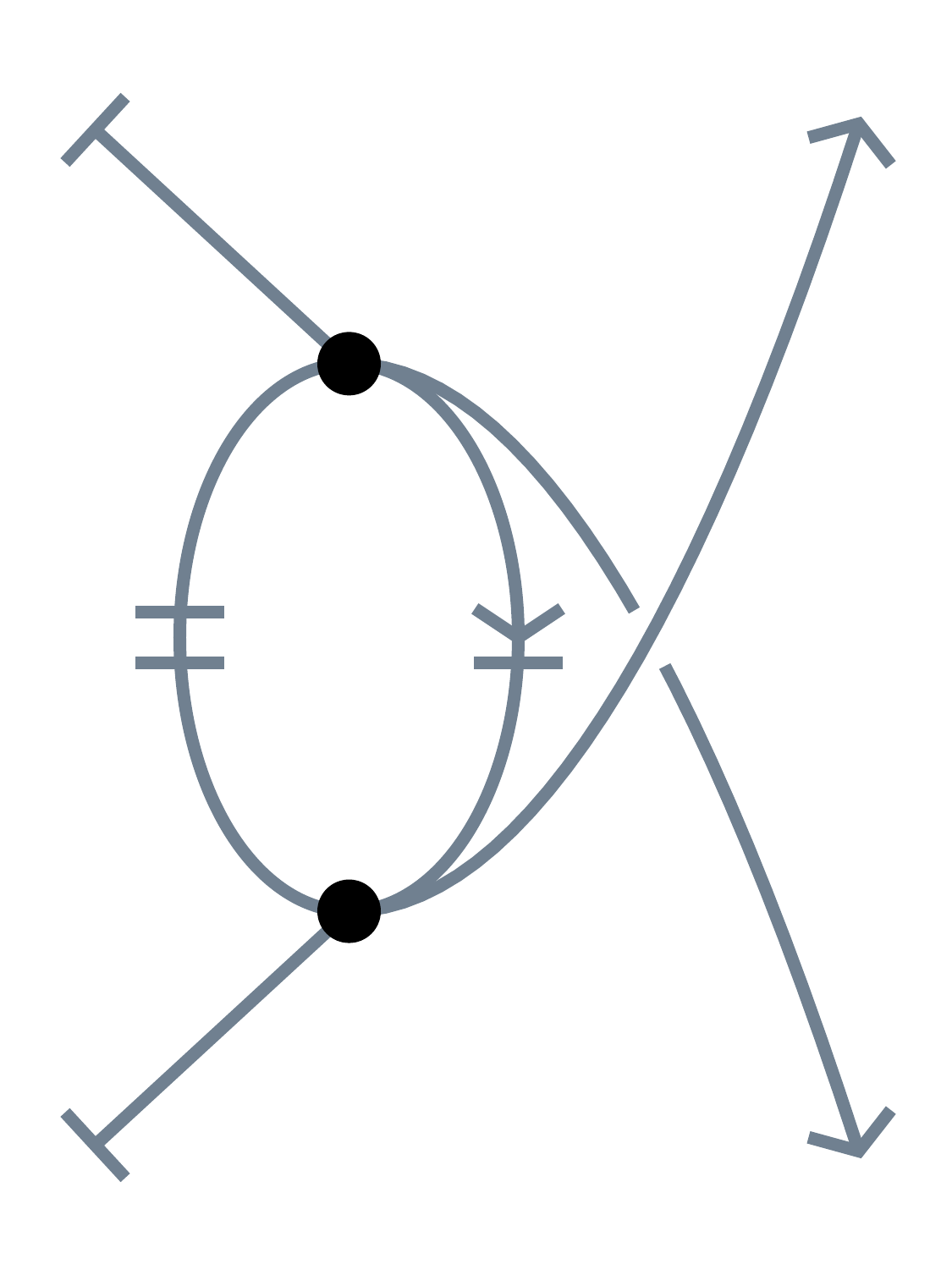}
			\subcaption{\label{fig_ccqq-u2}}}
	\end{minipage}
	\caption{\label{fig_ccqq}One-loop corrections for the vertex $\left(\phi^{\alpha}_c \phi^{\alpha}_q\right)^2$.\hfill\hfill}
\end{figure}

We note that the most advantage of working in the Keldysh basis is that not all loops are divergent. The only one-vertex and two-vertex loops that diverge near $d=4-\epsilon$ are shown in Fig.~\ref{fig_loop}, respectively. The first loop [Fig.~\ref{fig_loop1}] diverges as \[\overset{\text{dim.~reg.}}{\sim}-\frac{m^2}{4\pi^2 \epsilon},\] where ${\text{dim.~reg.}}$ stands for dimensional regularization, and the second loop [Fig.~\ref{fig_loop2}] diverges as \[\overset{\text{dim.~reg.}}{\sim}-\frac{i}{8\pi^2 \epsilon}.\]
Note that they are different from the results in Ref.~\cite{keldysh-renorm_bjlr17} by a factor of $2$, because in Ref.~\cite{keldysh-renorm_bjlr17} the Keldysh basis was defined differently from ours.

As a demonstration, here we explicitly calculate the second loop [Fig.~\ref{fig_loop2}] at $m^2= 0$. 
Introducing external legs with total momentum $p$ into the loop,
we have
\begin{eqnarray}
\label{eq_nscalar-loop2}
&&\int \frac{d^d k}{\left(2\pi\right)^d} \mathcal{P}^{R}(k)\mathcal{P}^{K}(p-k)\nonumber\\
&=&\int \frac{d^d k}{\left(2\pi\right)^d}\frac{i}{\left(\omega+ i\varepsilon\right)^2-\mathbf{k}^2} 2\pi \delta (\left(p-k\right)^2)\nonumber\\
&=&-i\int \frac{d^d k_\text{E}}{\left(2\pi\right)^d}\frac{1}{k_\text{E}^2} \frac{1}{\left(p_\text{E}-k_\text{E}\right)^2}\nonumber\\
&=&-i \frac{\Gamma(2-d/2)}{2^d \pi^{d/2}}\left(p_\text{E}^2\right)^{d/2-2}\int_0^1 dx \left[x\left(1-x\right)\right]^{d/2-2}\nonumber\\
&=& i C_{\epsilon}^{-1} \left(-p^2\right)^{-\epsilon/2}\sim -{i}/{\left(8\pi^2 \epsilon\right)}
\end{eqnarray}
where $C_{\epsilon}=-2^{5-2\epsilon}\pi^{\left(1-\epsilon\right)/2}\Gamma(3/2-{\epsilon}/{2}) \sin({\pi \epsilon}/{2})$. In the third step, we have used both the FDR for Eq.~(\ref{eq_nscalar-propagator}) and the momentum-space Wick rotation $\int {d^d k} \mathcal{P}(k^2)=i \int {d^d k_\text{E}} \mathcal{P}(-k_\text{E}^2)$.

A direct by-product of Eq.~(\ref{eq_nscalar-loop2}) is the two-point function between the two composite operators,  $\mathcal{O}_c=\phi^{\alpha}_c \phi^{\alpha}_c/\sqrt{2}$ and $\mathcal{O}_q=\sqrt{2}\phi^{\alpha}_c \phi^{\alpha}_q$, which can be calculated by
\begin{eqnarray*}
&&\left\langle\phi^{\alpha}_c\phi^{\alpha}_c(x)\phi^{\beta}_c\phi^{\beta}_q(0)\right\rangle\nonumber\\
&=& 2N\left[\int \frac{d^d k}{\left(2\pi\right)^d} \mathcal{P}^{R}(k)e^{-ikx}\right]\left[\int \frac{d^d k}{\left(2\pi\right)^d} \mathcal{P}^{K}(k)e^{-ikx}\right]\nonumber\\
&=& 2N \int \frac{d^d k}{\left(2\pi\right)^d} \int \frac{d^d p}{\left(2\pi\right)^d} \mathcal{P}^{R}(p)\mathcal{P}^{K}(k-p)e^{-ikx}\nonumber\\
&=&  \int \frac{d^d k}{\left(2\pi\right)^d} \left(2N i C_{\epsilon}^{-1}\right)\left[\mathbf{k}^2-\left(\omega+ i\varepsilon\right)^2\right]^{-\epsilon/2}e^{-ikx}.
\end{eqnarray*}
It is clear that the interior of this integral is the retarded function $G^{R}_\text{Gauss}$ at the Gaussian fixed point for $\mathcal{O}$.

Below, we calculate the renormalization flow by looking at the perturbation of each vertex in details. 

\subsection{${\phi^{\alpha}_c\phi^{\alpha}_c\phi^{\beta}_c\phi^{\beta}_q}$}

Combining the diagrams above, now we look at the perturbative corrections for the vertex ${\phi^{\alpha}_c\phi^{\alpha}_c\phi^{\beta}_c\phi^{\beta}_q}$, which are given by three diagrams (s, t, and u channels) in total at the one-loop level, as shown in Fig.~\ref{fig_cccq}. The full correction to $\lambda$ is
\[
\left[-\frac{i}{2\cdot 3!}\left(3!\right)\right]^{-1}
\frac{1}{2}\left(-\frac{i \lambda}{2\cdot 3!}\right)^2\cdot 3\left[72+8\left(N-1\right)\right]\cdot\frac{-i}{8\pi^2 \epsilon},
\]
from which $\beta_{\bar{\lambda}}$ is derived and is given in the main text.

\subsection{$\left(\phi^{\alpha}_c \phi^{\alpha}_q\right)^2$}

Similarly, for $\left(\phi^{\alpha}_c \phi^{\alpha}_q\right)^2$, there are five diagrams in total (Fig.~\ref{fig_ccqq}) that contribute. The full correction to $\gamma$ is
\begin{eqnarray*}
\left[-2\left(2!2!\right)\right]^{-1}
\left(-2\gamma\right)\left(-\frac{i \lambda}{2\cdot 3!}\right)\nonumber\\
\cdot \left\{24+4\left[24+4\left(N-1\right)\right]\right\}\cdot\frac{-i}{8\pi^2 \epsilon},
\end{eqnarray*}
from which $\beta_{\bar{\gamma}}$ is derived and is given in the main text.

\begin{figure}[t!]
	{\includegraphics[width=40mm]{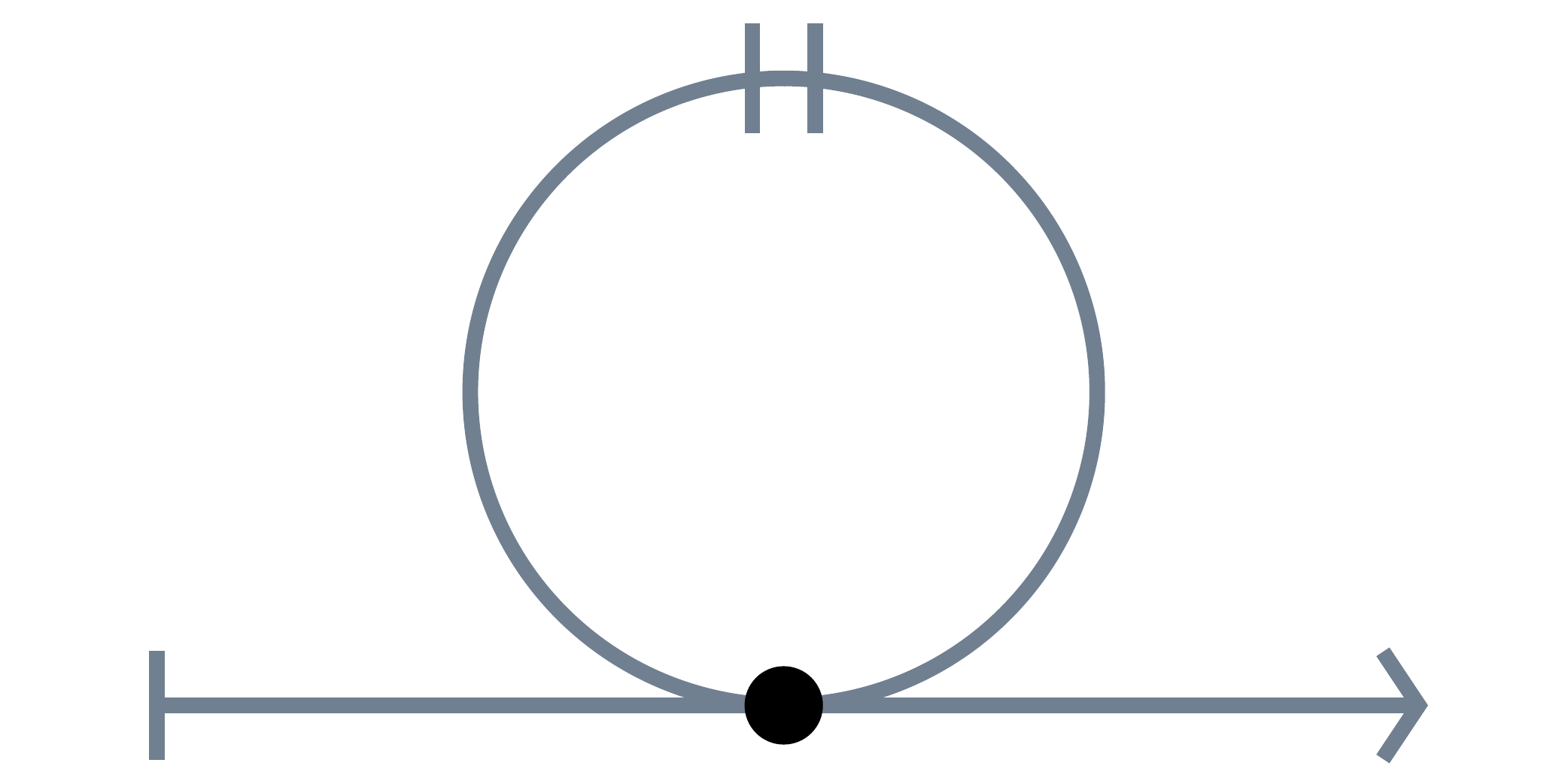}}
	\caption{\label{fig_cq}One-loop correction for the vertex $\phi^{\alpha}_c \phi^{\alpha}_q$.\hfill\hfill}
\end{figure}
\begin{figure}[t!]
	\centering	
	{\includegraphics[width=40mm]{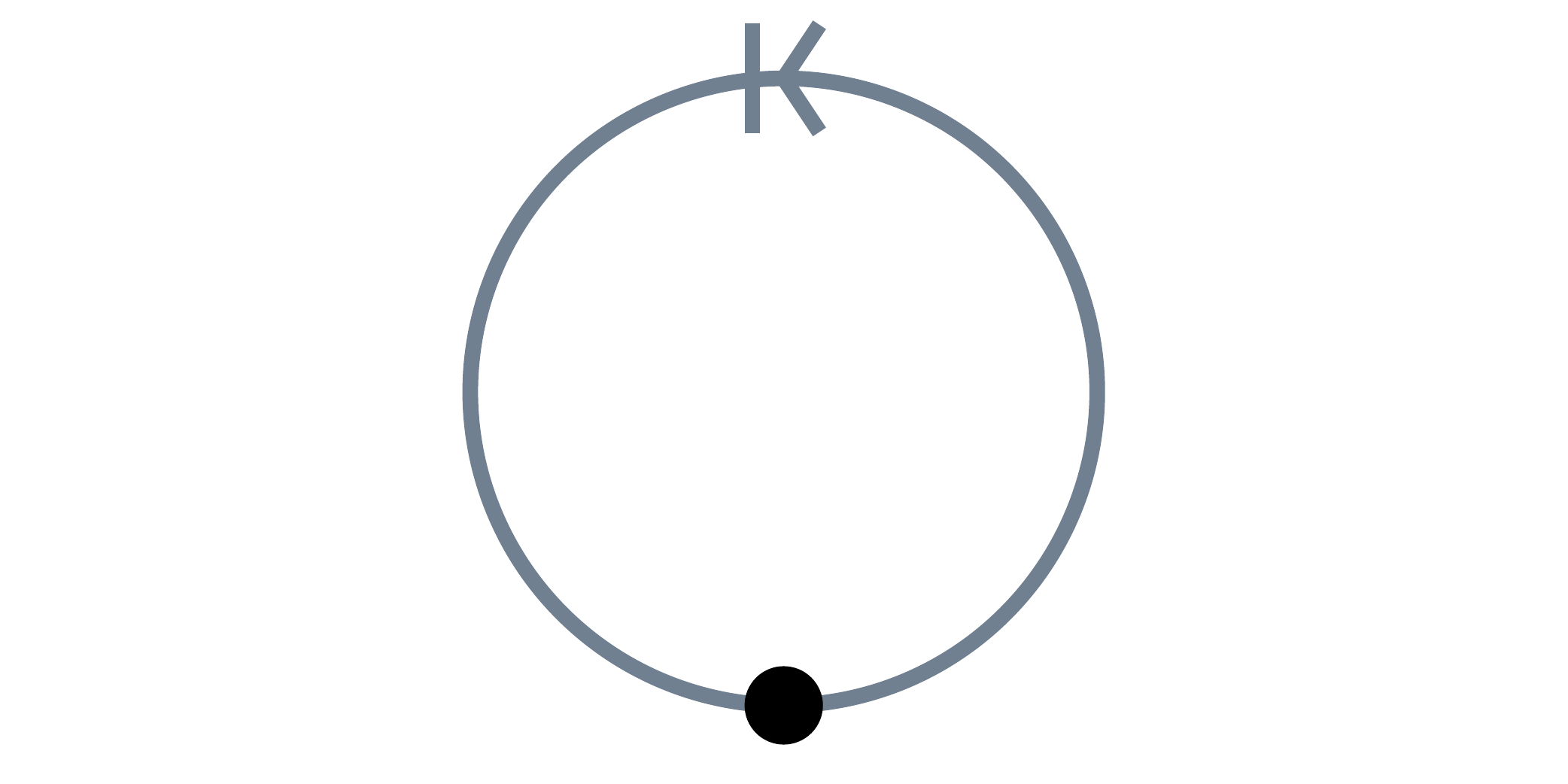}}
	\caption{\label{fig_loop3}New divergent loop in the Keldysh basis at $O(c^1)$.\hfill\hfill}
\end{figure}
\begin{figure}[t!]
	{\includegraphics[width=40mm]{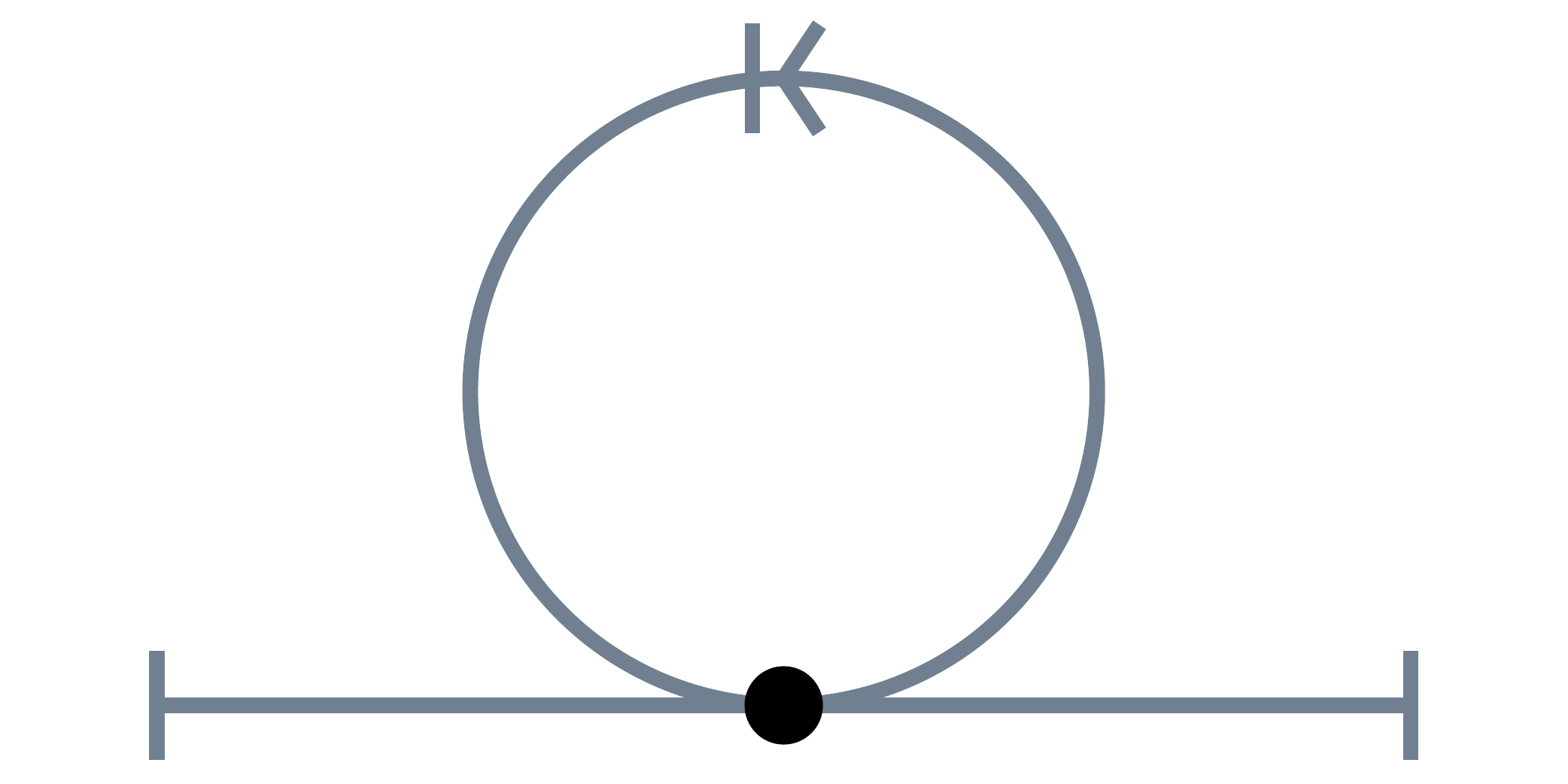}}
	\caption{\label{fig_cc}One-loop correction for the vertex $\phi^{\alpha}_c \phi^{\alpha}_c$.\hfill\hfill}
\end{figure}

\subsection{$\phi^{\alpha}_c \phi^{\alpha}_q$}

For $\phi^{\alpha}_c \phi^{\alpha}_q$, there is only one diagram (Fig.~\ref{fig_cq}). The full correction to $m^2$ is
\[
\left(-i\right)^{-1}
\left(-\frac{i \lambda}{2\cdot 3!}\right)\cdot \left[3+\left(N-1\right)\right]\cdot\frac{-m^2}{4\pi^2 \epsilon},
\]
from which $\beta_{\bar{m}^2}$ is derived.

\subsection{$\phi^{\alpha}_c \phi^{\alpha}_c$}

The vertex $\phi^{\alpha}_c \phi^{\alpha}_c$ needs special treatment, as we have to let $c$ be nonzero and consider the perturbative correction to the free propagators $\boldsymbol{\mathcal{P}}$.
In particular, the new perturbed retarded/advanced propagators are not $\mathcal{P}^{R/A}$, but $\mathcal{P}^{R/A}-c\mathcal{P}^{R/A}\mathcal{P}^{K}+O(c^2)$, the second term of which now diverges in the same way as Fig.~\ref{fig_loop2} does. 

Up to $O(c^1)$, we also have to include a new one-vertex loop (Fig.~\ref{fig_loop3}) which diverges as \[\overset{\text{dim.~reg.}}{\sim}\frac{ic}{8\pi^2 \epsilon}.\]
Therefore, taken into account the new diagram (Fig.~\ref{fig_cc}) at $O(c^1)$, the full correction to $c$ is
\[
\left[-\frac{1}{2}\left(2!\right)\right]^{-1}
\left(-\frac{i \lambda}{2\cdot 3!}\right)\cdot \left[6+2\left(N-1\right)\right]\cdot\frac{ic}{8\pi^2 \epsilon},
\]
from which $\beta_{\bar{c}}$ is derived.

\newpage
\bibliography{OQFTConformDeform}

\end{document}